\documentclass{aa}
\usepackage{graphicx}
\usepackage{natbib,twoopt}
\usepackage{flushend}
\atColsBreak{\vskip+4pt}
\usepackage[colorlinks,linkcolor=blue,urlcolor=blue,citecolor=black]{hyperref} %% to avoid \citeads line fills
\bibpunct{(}{)}{;}{a}{}{,}             %% natbib format for A&A and ApJ
\makeatletter
  \newcommandtwoopt{\citeads}[3][][]{\href{http://ui.adsabs.harvard.edu/abs/#3}%
    {\def\hyper@linkstart##1##2{}%
     \let\hyper@linkend\@empty\citealp[#1][#2]{#3}}}
  \newcommandtwoopt{\citepads}[3][][]{\href{http://ui.adsabs.harvard.edu/abs/#3}%
    {\def\hyper@linkstart##1##2{}%
     \let\hyper@linkend\@empty\citep[#1][#2]{#3}}}
  \newcommandtwoopt{\citetads}[3][][]{\href{http://ui.adsabs.harvard.edu/abs/#3}%
    {\def\hyper@linkstart##1##2{}%
     \let\hyper@linkend\@empty\citet[#1][#2]{#3}}}
   \newcommandtwoopt{\citeauthorads}[3][][]%
    {\href{http://ui.adsabs.harvard.edu/abs/#3}
    {\def\hyper@linkstart##1##2{}%
     \let\hyper@linkend\@empty\citeauthor[#1][#2]{#3}}}
  \newcommandtwoopt{\citeyearads}[3][][]%
    {\href{http://ui.adsabs.harvard.edu/abs/#3}
    {\def\hyper@linkstart##1##2{}%
     \let\hyper@linkend\@empty\citeyear[#1][#2]{#3}}}
  \renewcommand*\aa@pageof{, page \thepage{} of \pageref*{LastPage}} % to get rid of "Package hyperref Warning: Suppressing link with empty target on input line 92." warnings
\makeatother
\usepackage[varg]{txfonts}
\usepackage[font=small,labelfont=bf,labelsep=period]{caption} % for multi-page figures with \ContinuedFloat
\begin{document}

%\titlerunning{}
\title{A proto-helium white dwarf stripped by a substellar companion via common-envelope ejection}
\subtitle{Uncovering the true nature of a candidate hypervelocity B-type star%
\thanks{The fully reduced and wavelength calibrated ESI and ISIS spectra are only available at the CDS via anonymous ftp to \url{cdsarc.u-strasbg.fr (130.79.128.5)} or via \url{http://cdsarc.u-strasbg.fr/viz-bin/cat/J/A+A/650/A102}}
}
\author{A.~Irrgang\inst{\ref{remeis}}
        \and
        S.~Geier\inst{\ref{potsdam}}
        \and
        U.~Heber\inst{\ref{remeis}}
        \and
        T.~Kupfer\inst{\ref{kavli}}
        \and
        K.~El-Badry\inst{\ref{berkeley}}
        \and
        S.~Bloemen\inst{\ref{nijmegen}}
       }

\institute{
Dr.~Karl~Remeis-Observatory \& ECAP, Astronomical Institute, Friedrich-Alexander University Erlangen-Nuremberg (FAU), Sternwartstr.~7, 96049 Bamberg, Germany\label{remeis}\\
\email{andreas.irrgang@fau.de}
\and
Institut f\"ur Physik und Astronomie, Universit\"at Potsdam, Karl-Liebknecht-Str.\ 24/25, 14476 Potsdam, Germany\label{potsdam}
\and
Kavli Institute for Theoretical Physics, University of California, Santa Barbara, CA 93106, USA\label{kavli}
\and
Department of Astronomy and Theoretical Astrophysics Center, University of California Berkeley, Berkeley, CA 94720, USA\label{berkeley}
\and
Department of Astrophysics/IMAPP, Radboud University Nijmegen, P.O. Box 9010, 6500 GL Nijmegen, The Netherlands\label{nijmegen}
}
\date{Received 25 June 2020 / Accepted 7 April 2021}

\abstract{In the past, SDSS\,J160429.12$+$100002.2 was spectroscopically classified as a blue horizontal branch (BHB) star. Assuming a luminosity that is characteristic of BHB stars, the object's radial velocity and proper motions from {\it Gaia} Early Data Release 3 would imply that its Galactic rest-frame velocity exceeds its local escape velocity. Consequently, the object would be considered a hypervelocity star, which would prove particularly interesting because its Galactic trajectory points in our direction. However, based on the spectroscopic analysis of follow-up observations, we show that the object is actually a short-period ($P \approx 3.4$\,h) single-lined spectroscopic binary system with a visible B-type star (effective temperature $T_{\mathrm{eff}} = 15\,840\pm160$\,K and surface gravity $\log(g) = 4.86\pm0.04$) that is less luminous than typical BHB stars. Accordingly, the distance of the system is lower than originally thought, which renders its Galactic orbit bound to the Galaxy. Nevertheless, it is still an extreme halo object on a highly retrograde orbit. The abundances of He, C, N, O, Ne, Mg, Al, Si, S, and Ca are subsolar by factors from 3 to more than 100, while Fe is enriched by a factor of about 6. This peculiar chemical composition pattern is most likely caused by atomic diffusion processes. Combining constraints from astrometry, orbital motion, photometry, and spectroscopy, we conclude that the visible component is an unevolved proto-helium white dwarf with a thin hydrogen envelope that was stripped by a substellar companion through common-envelope ejection. Its unique configuration renders the binary system an interesting test bed for stellar binary evolution in general and common-envelope evolution in particular.
}


\maketitle
\section{\label{sect:intro}Introduction}
The star SDSS\,J160429.12$+$100002.2 (J1604$+$1000 for short) is a rather faint ($G=17.13$\,mag) blue ($G_\mathrm{BP}-G_\mathrm{RP}=-0.16$\,mag) star of relatively high Galactic latitude ($b=+41.5$\,deg) that was discovered in a search for white dwarf (WD) stars in the tenth data release (DR10) of the Sloan Digital Sky Survey (SDSS; \citeads{2014ApJS..211...17A}) by \citetads{2015MNRAS.448.2260G} and \citetads{2015MNRAS.446.4078K}. While the former classify it as a ``narrow-line hydrogen star,'' that is, a star with a hydrogen-dominated atmosphere but a surface gravity that is lower than those of typical WDs, the latter report a WD spectral type of a class DAB star that is flagged as uncertain, with effective temperature $T_{\mathrm{eff}} = 23\,819\pm270$\,K and surface gravity $\log(g) = 5.68\pm 0.03$, which is indeed too low for a WD. Because these atmospheric parameters are typical of hot subdwarf stars (sdO/B), \citetads{2017A&A...600A..50G} assigned the star to the class of B-type subdwarfs (sdB).

J1604$+$1000 caught our attention because it exhibits a high negative radial velocity ($-345\pm3$\,km\,s${}^{-1}$; \citeads{2017A&A...600A..50G}). Consequently, the object was selected for spectroscopic follow-up observations in an ongoing project that is looking for hot subdwarfs with extreme radial velocities in order to find high-amplitude radial-velocity variables, that is, short-period close binaries (MUCHFUSS; \citeads{2015A&A...577A..26G}) as well as high-speed stars unbound to the Galaxy, such as the hypervelocity hot subdwarf star US\,708 (\citeads{2005A&A...444L..61H}, \citeads{2015Sci...347.1126G}). The initial survey, the Hyper-MUCHFUSS project, was restricted to sdO/B stars with radial velocities exceeding $\pm100$\,km\,s${}^{-1}$ (\citeads{2011A&A...527A.137T}, \citeads{2016ApJ...821L..13N}, \citeads{2017A&A...601A..58Z}).

Based on the SDSS spectrum, which was the only one available at the time of the aforementioned study, a preliminary spectroscopic analysis was carried out using a grid of synthetic spectra that is based on model atmospheres in local thermodynamic equilibrium (LTE) that account for metal-line blanketing \citepads{2000A&A...363..198H}. By means of a $\chi^2$ minimization technique \citepads{1999ApJ...517..399N}, best fitting atmospheric parameters ($T_{\mathrm{eff}} = 16\,800\pm500$\,K, $\log(g) = 4.72\pm0.09$, and $\log(n(\mathrm{He})/n(\mathrm{H}))=-2.2$) were found that are, on the one hand, considerably different from the ones derived by \citetads{2015MNRAS.446.4078K} and, on the other hand, consistent with those of blue horizontal branch (BHB) stars in globular clusters (see, e.g., \citeads{2007A&A...474..505M}; \citeads{2011A&A...526A.136M}).

Assuming a typical BHB mass of $0.5\pm0.1\,M_\odot$, those preliminary atmospheric parameters gave a spectrophotometric distance $d=6.0\pm0.9$\,kpc that, when combined with the ground-based proper motions available back then, resulted in such a high Galactic rest-frame velocity, $\varv_\mathrm{Grf}$, that the star seemed to be a hypervelocity star, that is, an object that is gravitationally unbound to our Galaxy. This picture did not change when improved proper motions from {\it Gaia} Early Data Release 3 (EDR3; \citeads{2020arXiv201201533G}; \citeads{2020arXiv201203380L}) became available later on, which yielded $\varv_\mathrm{Grf} = 740\pm140$\,km\,s$^{-1}$. Contrary to almost all known hypervelocity stars, the preliminary Galactic orbit of J1604$+$1000 pointed in our direction, which would hint at the origins of the star lying outside of our Galaxy if, indeed, it were to be a BHB star.

In the meantime, high-quality follow-up spectra were taken during the Hyper-MUCHFUSS project. The implications of the analysis of those spectra are presented here: We demonstrate that J1604$+$1000 is not a hypervelocity B-type star but instead a short-period single-lined spectroscopic binary system on a highly retrograde Galactic orbit that hosts the progenitor of a low-mass helium white dwarf (He WD) and an unseen companion that is most likely a brown dwarf (BD).
\section{Analysis}
This section presents the technical details of the spectroscopic analysis (Sect.~\ref{sect:spectroscopy}), the investigation of the spectral energy distribution (SED; Sect.~\ref{sect:photometry}), the light-curve analysis (Sect.~\ref{sect:lightcurve}), the modeling of the radial-velocity curve (Sect.~\ref{sect:radial_velocity_curve}), the Bayes\-ian inference of stellar parameters (Sect.~\ref{sect:Bayesian_inference}), and the kinematic evaluation (Sect.~\ref{sect:kinematic_analysis}).
\subsection{\label{sect:spectroscopy}Quantitative spectroscopic analysis}
\begin{table}
\small
\centering
\setlength{\tabcolsep}{0.145cm}
\caption{\label{table:list_of_observations}Observations.}
\begin{tabular}{ccccc}
\hline\hline
Date & Exp. & S/N & Spectrograph & $\varv_{\mathrm{rad}}$\,\tablefootmark{(a)} \\
(d) & (s) & & & (km\,s${}^{-1}$) \\
\hline
$5706.8079$ & $6306$ & $36$ & BOSS (2000) & $-320.6\pm4.0$\,\tablefootmark{(b)} \\
$7579.8213$ & $1800$ & $43$ & ESI (5200) & $-383.8\pm3.3$ \\
$7579.8427$ & $1800$ & $12$ & ESI (5200) & $-370.5\pm4.4$ \\
$7579.8654$ & $1800$ & $47$ & ESI (8000) & $-343.3\pm3.2$ \\
$7959.3798$ &  $900$ & $35$ & blue arm ISIS (2.0\,\AA) & $-315.2\pm11.6$\,\tablefootmark{(c)} \\
$7959.3908$ &  $900$ & $38$ & blue arm ISIS (2.0\,\AA) & $-308.0\pm11.5$\,\tablefootmark{(c)} \\
$7961.4907$ & $1800$ & $34$ & blue arm ISIS (1.2\,\AA) & $-378.6\pm11.0$\,\tablefootmark{(c)} \\
$7961.5132$ & $1800$ & $27$ & blue arm ISIS (1.2\,\AA) & $-323.5\pm11.2$\,\tablefootmark{(c)} \\
$8255.0640$ & $1500$ & $62$ & ESI (8000) & $-343.5\pm6.8$\,\tablefootmark{(d)} \\
$8255.0821$ & $1500$ & $60$ & ESI (8000) & $-369.0\pm5.3$\,\tablefootmark{(d)} \\
$8564.7278$ & $1800$ & $19$ & X-shooter (10\,000) & $-323.4\pm4.0$ \\
$8565.7407$ & $1800$ & $28$ & X-shooter (10\,000) & $-337.9\pm3.2$ \\
$8565.7668$ & $1800$ & $35$ & X-shooter (10\,000) & $-375.8\pm3.2$ \\
$8565.8164$ & $1800$ & $30$ & X-shooter (10\,000) & $-359.3\pm4.4$ \\
$8565.8427$ & $1800$ & $36$ & X-shooter (10\,000) & $-317.4\pm3.4$ \\
$8565.8693$ & $1800$ & $36$ & X-shooter (10\,000) & $-318.5\pm4.0$ \\
\hline
\end{tabular}
\tablefoot{The first column is the heliocentric Julian date (HJD) at the middle of the observation minus 2\,450\,000, the second column is the exposure time, the third is the average signal-to-noise ratio (S/N), the fourth is the spectrograph with its approximate resolving power $\lambda/\Delta\lambda$ or $\Delta\lambda$ in parentheses, and the fifth is the measured heliocentric radial velocity. The uncertainties of the last column's values are the quadratic sum of $1\sigma$ statistical uncertainties and a generic systematic uncertainty of 3\,km\,s$^{-1}$. \tablefoottext{a}{Not corrected for gravitational redshift.} \tablefoottext{b}{Not used for the analysis of the radial-velocity curve because the exposure time is roughly half the orbital period.} \tablefoottext{c}{A generic uncertainty of 10\,km\,s$^{-1}$ was added in quadrature to account for possible wavelength shifts caused by instrument flexure.} \tablefoottext{d}{Uncertainties were increased to account for a small offset in the wavelength calibration as indicated by the position of the telluric features.}}
\end{table}
\begin{table*}
\small
\centering
\renewcommand{\arraystretch}{1.2}
\setlength{\tabcolsep}{0.083cm}
\caption{\label{table:atmospheric_parameters} Atmospheric parameters and elemental abundances.}
\begin{tabular}{lrrrrrrrrrrrrrrrrrr}
\hline\hline
 & $T_{\mathrm{eff}}$ & $\log(g)$ & $\varv\sin(i)$ & $\xi$ & & \multicolumn{12}{c}{$\log(n(x))$} \\
\cline{4-5} \cline{7-18}
& (K) & (cgs) & \multicolumn{2}{c}{(km\,s$^{-1}$)} & & He & C & N & O & Ne & Mg & Al & Si & S & Ar & Ca & Fe\\
\hline
Value & $15\,840$ & $4.86$ & $18.4$\,\tablefootmark{(a)} & $2.0$ & & $-2.13$ & $\leq -5.66$ & $\leq -4.69$ & $\leq -5.36$ & $\leq -5.97$ & $-5.71$ & $\leq -7.08$ & $\leq -7.00$ & $\leq -6.60$ & $\leq -5.75$ & $-6.38$\,\tablefootmark{(b)} & $-3.69$
\\ Stat. & $^{+20}_{-20}$ & $^{+0.01}_{-0.01}$ & $^{+0.9}_{-1.0}$ & $^{+0.1}_{-0.1}$ & & $^{+0.02}_{-0.02}$ & \ldots & \ldots & \ldots & \ldots & $^{+0.04}_{-0.04}$ & \ldots & \ldots & \ldots & \ldots & $^{+0.07}_{-0.08}$ & $^{+0.02}_{-0.02}$
\\ Sys. & $^{+160}_{-160}$ & $^{+0.04}_{-0.04}$ & $^{+2.0}_{-1.7}$ & $^{+1.3}_{-1.5}$ & & $^{+0.04}_{-0.04}$ & \ldots & \ldots & \ldots & \ldots & $^{+0.04}_{-0.05}$ & \ldots & \ldots & \ldots & \ldots & $^{+0.20}_{-0.22}$ & $^{+0.09}_{-0.09}$ \\
\hline
$\odot$\,\tablefootmark{(c)} & & & & & & $-1.06$ & $-3.57$ & $-4.17$ & $-3.31$ & $-4.07$ & $-4.40$ & $-5.55$ & $-4.49$ & $-4.88$ & $-5.60$ & $-5.66$ & $-4.50$ \\
\hline
Value & & & & & & $-1.54$ & $\leq -4.60$ & $\leq -3.57$ & $\leq -4.18$ & $\leq -4.69$ & $-4.34$ & $\leq -5.67$ & $\leq -5.57$ & $\leq -5.11$ & $\leq -4.17$ & $-4.80$\,\tablefootmark{(b)} & $-1.96$
\\ Stat. & & & & & & $^{+0.02}_{-0.02}$ & \ldots & \ldots & \ldots & \ldots & $^{+0.04}_{-0.04}$ & \ldots & \ldots & \ldots & \ldots & $^{+0.07}_{-0.08}$ & $^{+0.02}_{-0.02}$
\\ Sys. & & & & & & $^{+0.04}_{-0.04}$ & \ldots & \ldots & \ldots & \ldots & $^{+0.04}_{-0.05}$ & \ldots & \ldots & \ldots & \ldots & $^{+0.20}_{-0.22}$& $^{+0.09}_{-0.09}$ \\
\hline
$\odot$\,\tablefootmark{(c)} & & & & & & $-0.57$ & $-2.60$ & $-3.13$ & $-2.22$ & $-2.87$ & $-3.12$ & $-4.23$ & $-3.15$ & $-3.48$ & $-4.11$ & $-4.17$ & $-2.86$ \\
\hline
\end{tabular}
\tablefoot{The abundance $n(x)$ is given either as a fractional particle number (\textit{upper four rows}) or a mass fraction (\textit{lower four rows}) of species $x$ with respect to all elements. Statistical uncertainties (\textit{``Stat.''}) are $1\sigma$ confidence limits based on $\chi^2$ statistics. Systematic uncertainties (\textit{``Sys.''}) cover only the effects induced by additional variations of $1\%$ in $T_{\mathrm{eff}}$ and $0.04$ in $\log(g)$ and are formally taken to be $1\sigma$ confidence limits (see \citeads{2014A&A...565A..63I} for details). Abundances without uncertainties are upper limits because the respective chemical elements do not exhibit spectral lines that are strong enough to be measured. \tablefoottext{a}{Affected by orbital smearing and thus overestimated (see Sect.~\ref{sect:radial_velocity_curve}).} \tablefoottext{b}{Abundance based on population numbers computed in LTE.} \tablefoottext{c}{Proto-solar nebula values from \citetads{2009ARA&A..47..481A} as reference.}}
\end{table*}
Our spectral investigation is based on data from four different instruments. The very first spectrum was taken with the spectrograph of the Baryon Oscillation Spectroscopic Survey (BOSS; \citeads{2013AJ....146...32S}) attached to the 2.5\,m telescope at Apache Point Observatory. Another five spectra were taken with the Echellette Spectrograph and Imager (ESI; \citeads{2002PASP..114..851S}) mounted at the Keck~II telescope, four with the blue arm of the Intermediate-dispersion Spectrograph and Imaging System\footnote{\url{http://www.ing.iac.es/astronomy/instruments/isis/}} (ISIS) at the \textit{William Herschel} Telescope, and six with the X-shooter \citepads{2011A&A...536A.105V} at the ESO Very Large Telescope. Table~\ref{table:list_of_observations} lists the date, exposure time, average signal-to-noise ratio (S/N), and measured radial velocity, $\varv_\mathrm{rad}$, of the individual observations.

Following the analysis strategy outlined in \citetads{2014A&A...565A..63I}, we simultaneously fit all those spectra over their entire spectral range to determine the atmospheric parameters and elemental abundances. The underlying synthetic models were computed using a series of three codes. The structure of the atmosphere was computed in LTE with {\sc Atlas12} \citepads{1996ASPC..108..160K}. Based on this atmosphere, population numbers in non-LTE were calculated with the {\sc Detail} code (\citeads{1981PhDT.......113G}; \citealt{detailsurface2}), which numerically solves the coupled radiative transfer and statistical equilibrium equations. Using the non-LTE occupation numbers from {\sc Detail} and more detailed line-broadening data as input, the emerging spectrum was eventually computed with the {\sc Surface} code (\citeads{1981PhDT.......113G}; \citealt{detailsurface2}). Recent updates to all three codes (see \citeads{2018A&A...615L...5I} for details) with respect to (i) non-LTE effects on the atmospheric structure, (ii) the implementation of the occupation probability formalism \citepads{1994A&A...282..151H} for hydrogen, and (iii) new Stark broadening tables for hydrogen \citepads{2009ApJ...696.1755T} and neutral helium \citepads{1997ApJS..108..559B} were also considered.

The results of the spectroscopic analysis are summarized in Table~\ref{table:atmospheric_parameters}. The derived effective temperature, $T_{\mathrm{eff}} = 15\,840\pm160$\,K, shows that the star is of spectral type B. As exemplified in Fig.~\ref{fig:spectra}, all available spectra only exhibit spectral lines of H, He, Mg, Ca, and Fe, which is quite uncommon for ``normal'' B-type stars. For reference, main sequence stars of similar temperature and solar chemical composition also show lines of C, N, O, Ne, Al, Si, S, and Ar. Owing to the absence of spectral features of those elements, we are only able to provide upper limits for the respective abundances. The measured abundances of He, Mg, Ca\footnote{The Ca abundance is based on population numbers in LTE.}, and Fe significantly differ from those of the Sun. In terms of number fractions, He and Mg are under-abundant by more than 1\,dex and Ca by 0.7\,dex, while Fe is enriched by about 0.8\,dex. Upper limits for the abundances of C, N, O, Ne, Al, Si, and S are also significantly below solar (from 0.5\,dex for N to 2.5 dex for Si). The only exception from this trend is Ar, for which the upper limit is close to solar. This peculiar abundance pattern clearly hints at ongoing diffusion processes. The inferred value for the microturbulence, $\xi = 2.0^{+1.3}_{-1.5}$\,km\,s${}^{-1}$, is rather inconspicuous. In contrast, the radial velocity, $\varv_\mathrm{rad}$, turns out to be variable on a timescale of hours (see Table~\ref{table:list_of_observations}). Consequently, and as discussed in Sect.~\ref{sect:radial_velocity_curve}, our value for the projected rotational velocity, $\varv\sin(i) = 18.4^{+2.2}_{-2.0}$\,km\,s${}^{-1}$, is overestimated due to the effect of orbital smearing.
\subsection{\label{sect:photometry}Analysis of the spectral energy distribution}
\begin{table}
\small
\centering
\renewcommand{\arraystretch}{1.2}
\caption{\label{table:photometry_results}Parameters derived from the analysis of the SED.}
\begin{tabular}{lr}
\hline\hline
Parameter & Value \\
\hline
Angular diameter $\log(\Theta\,\mathrm{(rad)})$ & $-11.418\pm0.006$ \\
Color excess $E(44-55)$ & $0.058\pm0.021$\,mag \\
Extinction parameter $R(55)$ (fixed) & $3.02$ \\
Effective temperature $T_{\mathrm{eff}}$ & $15\,800\pm700$\,K \\
Blackbody temperature $T_{\mathrm{bb}}$ & $2300^{+400}_{-600}$\,K \\
Blackbody surface ratio & $17^{+16}_{-\phantom{0}5}$ \\
\hline
\end{tabular}
\tablefoot{The given uncertainties are single-parameter $1\sigma$ confidence intervals based on $\chi^2$ statistics with a reduced $\chi^2$ at the best fit of $1.04$.}
\end{table}
\begin{figure}
\centering
\includegraphics[width=1\columnwidth]{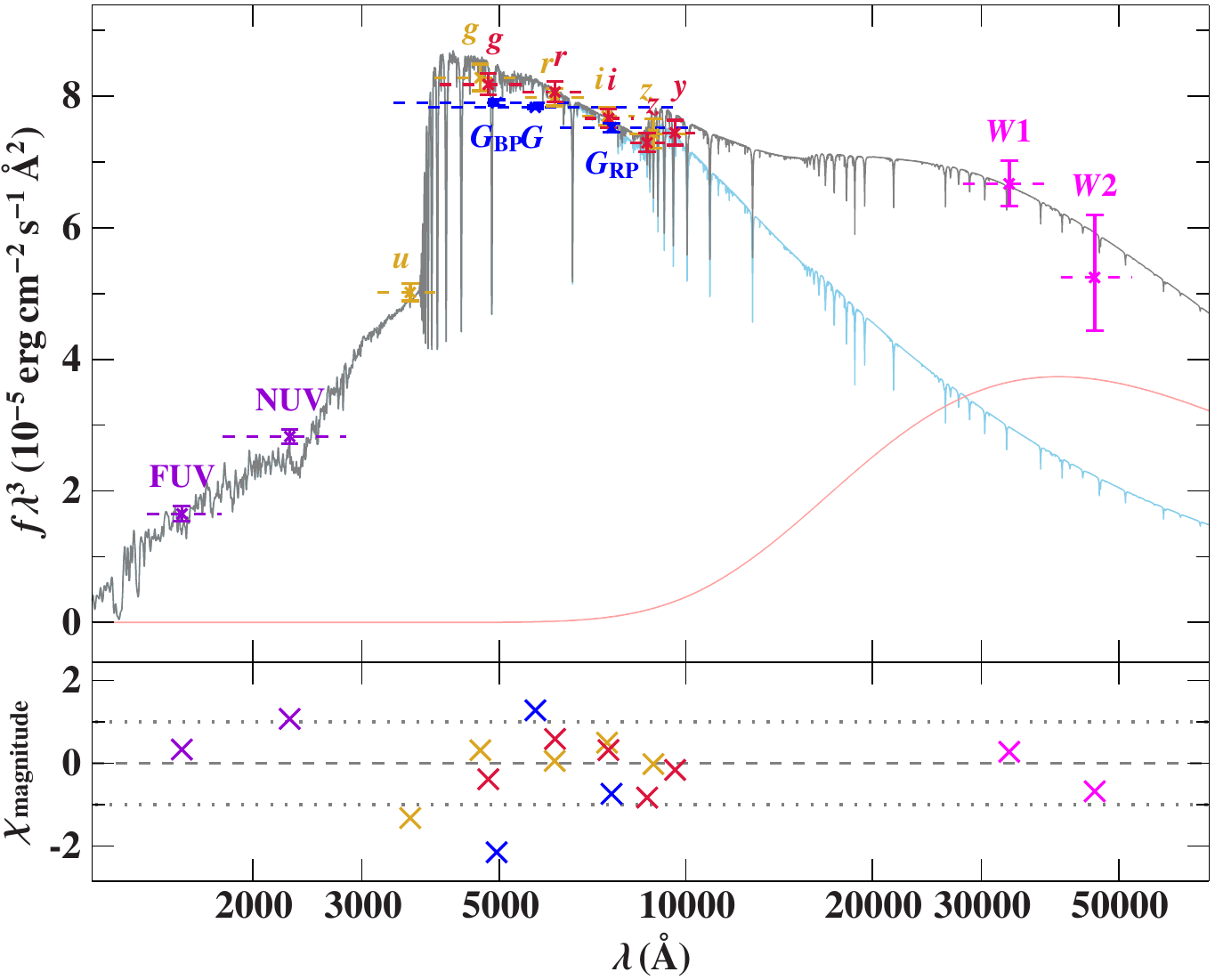}
\caption{\label{fig:photometry_sed}Comparison of synthetic and observed photometry. The \textit{top panel} shows the SED. The colored data points are filter-averaged fluxes that were converted from observed magnitudes (the respective full widths at tenth maximum of the filters are indicated by the dashed horizontal lines), while the solid gray line represents the best fitting model, i.e., it is based on the parameters from Table~\ref{table:photometry_results}, degraded to a spectral resolution of 6\,{\tiny\AA}. The flux is multiplied by the wavelength to the power of three to reduce the steep slope of the SED on such a wide wavelength range. The individual contributions of the stellar (light blue) and blackbody (light red) components are shown as well. The panel at the \textit{bottom} shows the residuals, $\chi$, that is, the difference between synthetic and observed magnitudes divided by the corresponding uncertainties. The photometric systems have the following color code: GALEX (violet; \citeads{2017yCat.2335....0B}), SDSS (golden; \citeads{2015ApJS..219...12A}), Pan-STARRS1 (red; \citeads{2017yCat.2349....0C}), {\it Gaia} (blue; \citeads{2020arXiv201201916R}), and WISE (magenta; \citeads{2019ApJS..240...30S}).}
\end{figure}
The SED provides an important constraint to cross-check spectroscopic results and to obtain a more comprehensive picture of the star. For this particular object, photometric measurements covering the ultraviolet, optical, and infrared are available. In order to validate our spectroscopic results, we fit the observed SED with synthetic SEDs computed with {\sc Atlas12}. Because the surface gravity, microturbulence, He abundance, and metallicity are, if at all, only poorly constrained by photometry, they were set to the values determined from spectroscopy (see Table~\ref{table:atmospheric_parameters}) using Fe as a proxy for metallicity. Consequently, the three quantities $T_{\mathrm{eff}}$, $\Theta$, and $E(44-55)$ remained as free parameters to match the observed SED. The angular diameter, $\Theta = 2 R_1/d$ ($R_1$ is the radius and $d$ the distance of the star; see \citeads{2018OAst...27...35H} for more details), was used as a distance scaling factor, and the color excess $E(44-55)$ was introduced to account for interstellar reddening, the effect of which was modeled here using the extinction law by \citetads{2019ApJ...886..108F} with a standard extinction coefficient of $R(55) = 3.02$. These extinction parameters are analogs of the more widely used color excess $E(B-V)$ and extinction $R(V)$ parameters, but with measurements in the Johnson $B$ and $V$ filters substituted by monochromatic ones at 4400\,\AA\ and 5500\,\AA\ (see, e.g., \citeads{2019ApJ...886..108F} for details). In order to empirically account for an apparent infrared excess appearing in the WISE data, we also modeled a blackbody component. This introduced two additional free yet highly anticorrelated parameters, namely a temperature and a flux weighting factor that was parameterized as a surface ratio relative to the stellar component. The results of the fitting procedure are summarized in Table~\ref{table:photometry_results} and illustrated in Fig.~\ref{fig:photometry_sed}. Effective temperatures from spectroscopy and photometry are basically identical, which corroborates our spectroscopic results. Moreover, the inferred interstellar reddening of $E(44-55) = 0.058\pm0.021$\,mag is consistent with upper limits from reddening maps (\citeads{1998ApJ...500..525S}: $E(B-V)=0.060\pm0.002$\,mag; \citeads{2011ApJ...737..103S}: $E(B-V)=0.051\pm0.002$\,mag). The best fitting parameters of the blackbody component are able to reproduce the observed infrared excess and hint at the presence of a very cool but relatively extended thermal source. Its effective radiation area is $17^{+16}_{-\phantom{0}5}$ times the projected surface area of the star itself.
\subsection{\label{sect:lightcurve}Analysis of the light curve}
\begin{figure}
\centering
\includegraphics[width=1\columnwidth]{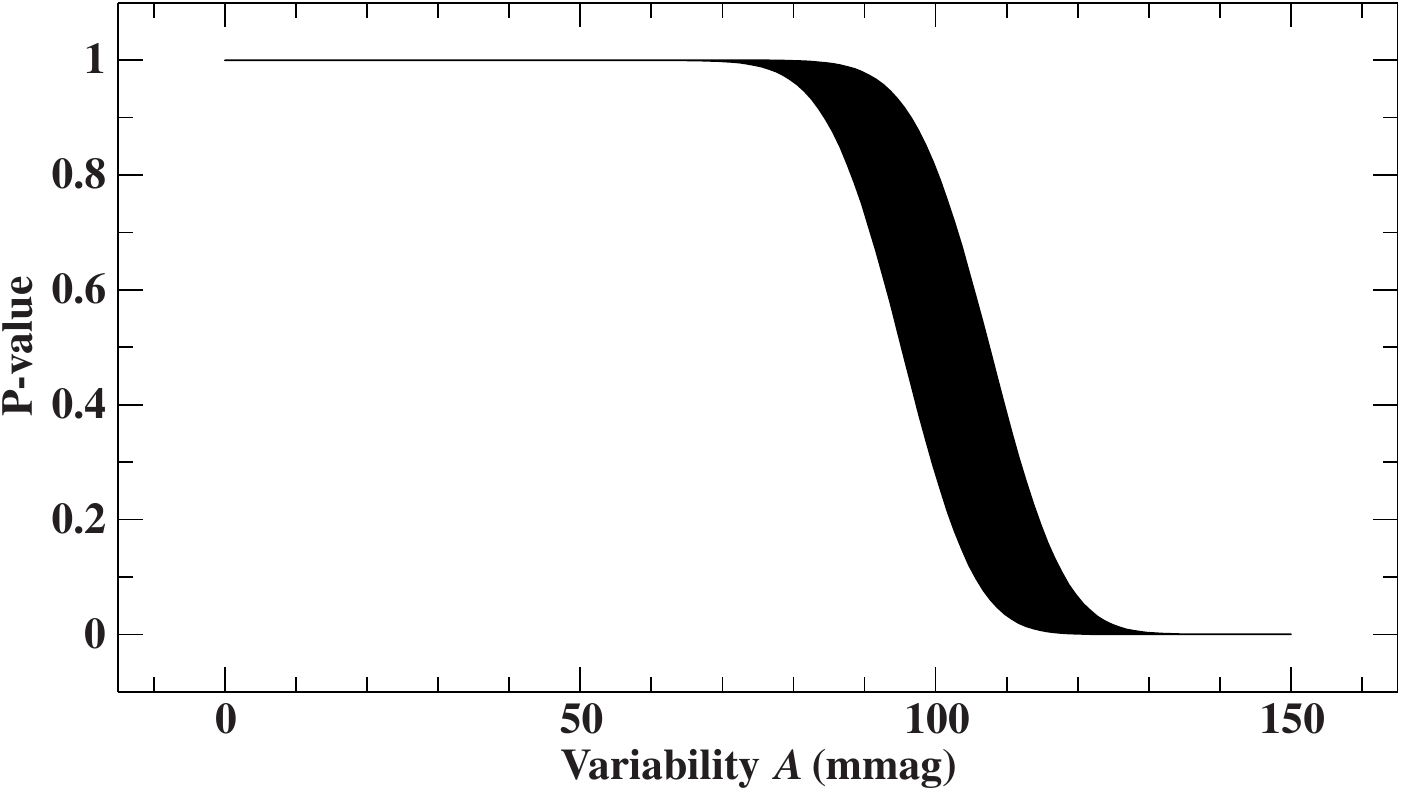}
\caption{\label{fig:lightcurve_pvalue}P-value inferred from $\chi^2$ testing the null hypothesis that the observed CSS light curve exhibits a certain amount of variability, $A$. The lower the p-value, the more unlikely this null hypothesis is. The width of the line represents $1\sigma$ uncertainties derived from modeling the unknown shape of the variability via a Monte Carlo simulation.}
\end{figure}
The Catalina Sky Survey (CSS; \citeads{2009ApJ...696..870D}) took a light curve of our target that consists of 379 $V$-band measurements spread over 3101 days. The mean magnitude of this data set is $V_\mathrm{CSS} = 17.11\pm0.01$\,mag, and the respective standard deviation is $0.08$\,mag. The latter is very close to the mean of the stated uncertainties on the individual measurements of $0.09$\,mag, which indicates that this light curve is heavily dominated by noise. Consequently, the available data set only allows an upper limit on the variability of the source to be determined. To this end, we employed a $\chi^2$ testing scheme to compute p-values for the null hypothesis that the source is variable with a certain amplitude, $A$. To be independent of any assumptions regarding the form of the underlying light curve, we created one million arbitrarily shaped mock light curves of the form
\begin{equation}%
V_{\mathrm{CSS},i} = V_\mathrm{CSS} + A X_i = 17.11\,\mathrm{mag} + A X_i
\label{eq:mock_lightcurve}
,\end{equation}
where $1 \leq i \leq 379$ is the index of a CSS data point and $X_{i}$ is a uniformly distributed random number between $-1$ and $1$. Starting from a constant source with $A=0$\,mag, this Monte Carlo procedure was repeated for steadily increasing values of $A$ until a p-value consistent with 0 was reached, that is, until it became completely unlikely to erroneously reject the null hypothesis despite it being true. The result of this exercise, which is illustrated in Fig.~\ref{fig:lightcurve_pvalue}, indicates that the available CSS light curve is consistent with the source being photometrically variable with a semi-amplitude of $\lesssim 110$\,mmag.
\subsection{\label{sect:radial_velocity_curve}Analysis of the radial-velocity curve}
Already in the first follow-up observation run in July 2016, variations in the radial velocity became obvious. Within about an hour, $\varv_\mathrm{rad}$ changed by more than 40\,km\,s${}^{-1}$. In a second run in April 2019, enough follow-up observations were collected to construct the radial-velocity curve. For the modeling of this curve, we did not consider the BOSS spectrum because its exposure time turned out to be roughly half the orbital period, which is too long to infer a representative radial velocity.

Given that it is much smaller than the gap between the observation blocks, the determination of the orbital period, $P$, was not trivial. To estimate an upper limit for $P$, we looked at the radial-velocity variations in the night on heliocentric Julian date (HJD) 2\,458\,565\, (see Table~\ref{table:list_of_observations}). Within 2.5\,h, $\varv_\mathrm{rad}$ increased from about $-376$\,km\,s$^{-1}$ to roughly $-317$\,km\,s$^{-1}$. This change in $\varv_\mathrm{rad}$ coincides fairly well with the maximum velocity amplitude inferred from all available observations. For a sinusoidal curve, it would take $1/2+N$ times the orbital period to accomplish this maximum change, where $N \geq 0$ is an integer. Consequently, $P \approx 2.5\,\mathrm{h}/(1/2+N)\lesssim5$\,h. To account for the various approximations in this reasoning, we adopted a conservative upper limit of $18$\,h. A lower limit for $P$ can be estimated by considering the effect of orbital smearing, that is, spectral lines are smeared out due to the change in radial velocity over the course of an exposure. Assuming that $\varv_\mathrm{rad}$ increases linearly from its minimum to its maximum value during half the orbital period, it follows that $(376-317)\,\mathrm{km\,s}{}^{-1}/(P/2)$ times the exposure times of those observations that have sufficient spectral resolution to resolve line broadening ($0.42$--$0.5$\,h) has to be lower than or equal to the measured value for the projected rotational velocity of $\varv\sin(i) = 18.4^{+2.2}_{-2.0}$\,km\,s${}^{-1}$, which yields $P \geq 0.42\,\mathrm{h}\times2\times59/20.6 = 2.4$\,h.
\begin{figure}
\centering
\includegraphics[width=1\columnwidth]{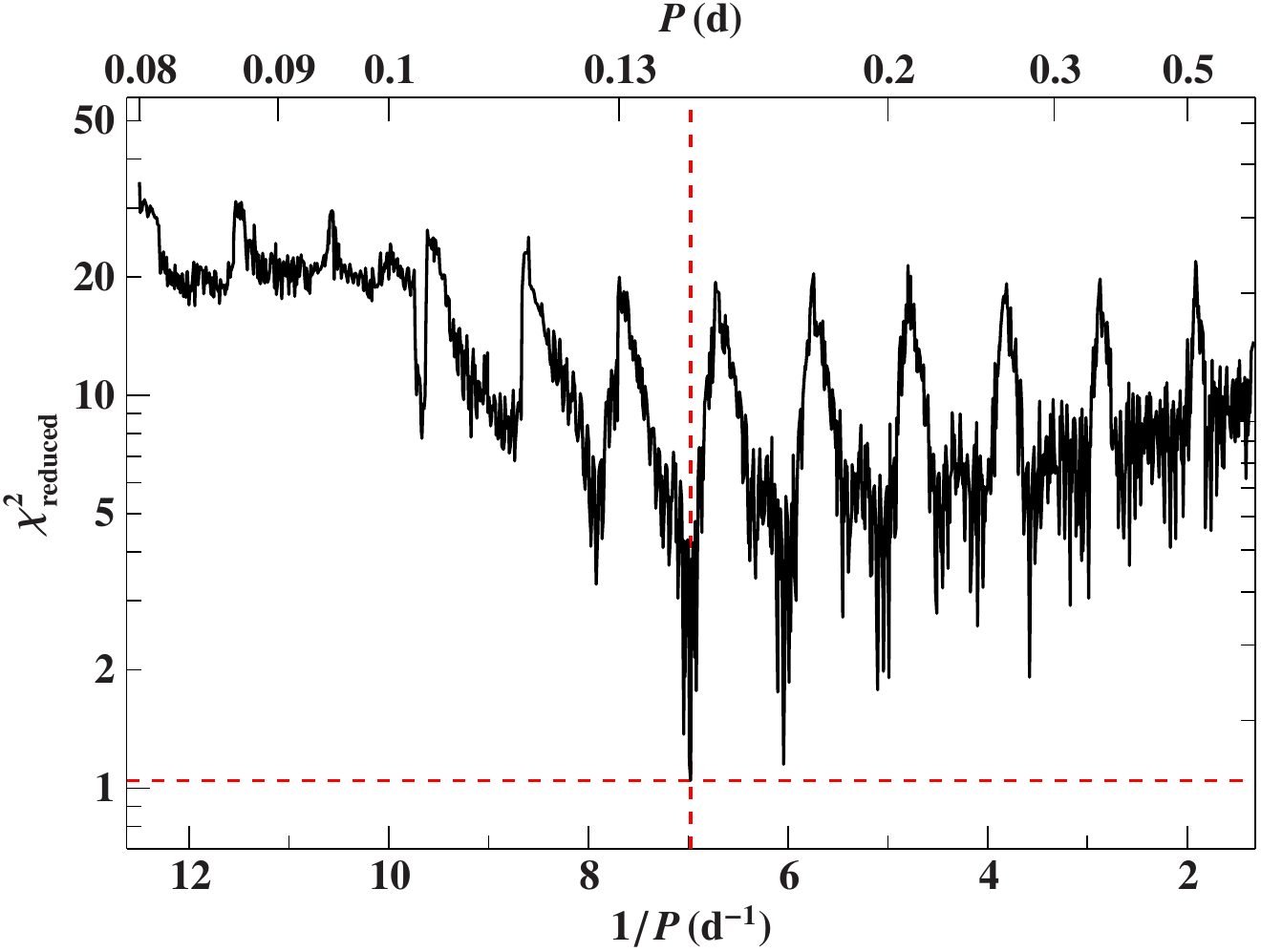}
\caption{\label{fig:periodogram}$\chi^2$ landscape (``periodogram''), which results from fitting the measured radial velocities with a Keplerian model, as a function of the orbital frequency or period. The step size in the orbital frequency was chosen such that phase shifts are always less than $0.01$. A minimum filter has been applied to lower the number of points in the plot to a reasonable amount. The dashed red lines mark the best fit.}
\end{figure}
\begin{table}
\small
\centering
\renewcommand{\arraystretch}{1.15}
\caption{\label{table:orbital_params}Orbital parameters.}
\begin{tabular}{lr}
\hline\hline
Parameter & Value \\
\hline
Period $P$ & $0.1433707\pm0.0000005$\,d \\
Epoch of periastron $T_{\mathrm{periastron}}$ & $2\,455\,705.89\pm0.15$\,HJD \\
Eccentricity $e$ & $0.04^{+0.07}_{-0.04}$ \\
Longitude of periastron $\omega$ & $350\pm180$\,deg \\
Velocity semi-amplitude $K_1$ & $37.6\pm2.0$\,km\,s${}^{-1}$ \\
Systemic velocity $\gamma$ & $-348.3\pm1.4$\,km\,s${}^{-1}$\,\tablefootmark{(a)} \\
\hline
Derived parameter & Value \\
\hline
Mass function $f$ & $(7.8\pm1.3)\times10^{-4}$\,$M_\odot$ \\
Projected semimajor axis $a_1 \sin(i)$ & $0.106\pm0.006$\,$R_\odot$ \\
Pericenter distance: $(1-e)\,a_1 \sin(i_\mathrm{o})$ & $0.102\pm0.008$\,$R_\odot$ \\
Apocenter distance: $(1+e)\,a_1 \sin(i_\mathrm{o})$ & $0.111\pm0.011$\,$R_\odot$ \\
\hline
\end{tabular}
\tablefoot{The given uncertainties are single-parameter $1\sigma$ confidence intervals based on $\chi^2$ statistics around the best fit with a reduced $\chi^2$ of 1.05. Except for $P$, the parameter values of the local minima in the periodogram (see Fig.~\ref{fig:periodogram}) are also covered by the given uncertainties. The quantities $T_{\mathrm{periastron}}$ and $\omega$ become degenerate for low-eccentricity orbits, which is why they are basically unconstrained here. \tablefoottext{a}{Not corrected for gravitational redshift.}}
\end{table}
\begin{figure}
\centering
\includegraphics[width=1\columnwidth]{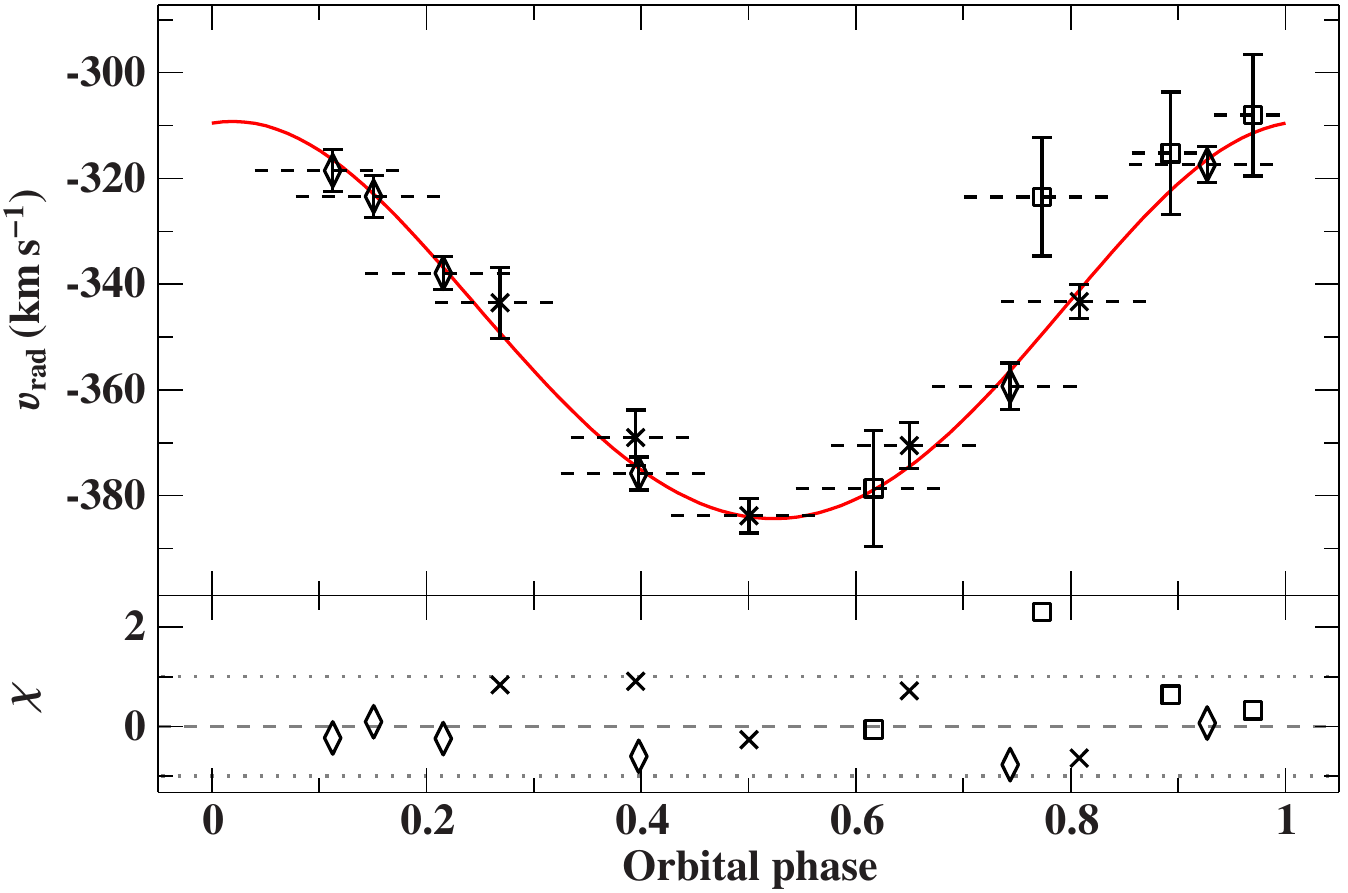}
\caption{\label{fig:vrad_curve}Phased radial-velocity curve. The measurements are represented by black symbols with $1\sigma$ error bars, and the best fitting Keplerian model is indicated by the solid red curve. The corresponding orbital parameters are given in Table~\ref{table:orbital_params}. The ESI spectra are shown as crosses, ISIS spectra as open squares, and X-shooter spectra as open diamonds. The dashed horizontal lines indicate the individual exposure times over which the Keplerian curve is averaged before being compared with the measurements. The residuals, $\chi$, are shown in the lower panel. An orbital phase of 0 corresponds to the pericenter passage.}
\end{figure}

Figure~\ref{fig:periodogram} shows the $\chi^2$ landscape that results from fitting the observed radial velocities with a Keplerian curve for orbital periods between the aforementioned limits. Instead of simply evaluating the Keplerian curve at the midpoint of the exposure times, we averaged it over the exposure times to account for the fact that they are non-negligible fractions of the orbital period. The parameters of the best fitting configuration are listed in Table~\ref{table:orbital_params}, and the corresponding phased radial-velocity curve is shown in Fig.~\ref{fig:vrad_curve}. The star is part of a very close, short-period single-lined spectroscopic binary system with a large negative systemic radial velocity. Although a non-zero eccentricity, $e$, is formally not excluded by the current data situation, a circularized orbit is most likely from an evolutionary point of view (see Sect.~\ref{sect:thin_hydrogen_envelope}). A more precise determination of $e$ requires follow-up observations that cover a few consecutive orbits with a single spectrograph in order to avoid the main sources of uncertainty that this study might be suffering.

With the orbital parameters from Table~\ref{table:orbital_params}, we can now turn the tables and check to what extent our estimate for the projected rotational velocity is actually affected by orbital smearing. Assuming again that $\varv_\mathrm{rad}$ increases linearly from its minimum to its maximum value during half the orbital period, the amount of orbital smearing can be estimated to be $4 \times K_1 / P \approx 44$\,km\,s${}^{-1}$\,h${}^{-1}$ times the relevant exposure times ($0.42$--$0.5$\,h), which yields values between $18$--$22$\,km\,s${}^{-1}$. Consequently, the measured projected rotational velocity of $\varv\sin(i) = 18.4^{+2.2}_{-2.0}$\,km\,s${}^{-1}$ is heavily affected by orbital smearing and its true value is probably much lower, which implies that the object is a slow rotator or that we see the star's rotational axis under a very small inclination, $i$.
\subsection{\label{sect:Bayesian_inference}Bayesian inference of stellar parameters}
Based on Bayesian inference, the radius and mass of the visible component ($R_1$, $M_1$), the mass of the unseen companion ($M_2$), and the orbital inclination ($i_\mathrm{o}$), which need not be the same as the star's inclination ($i$), can be constrained from the measured angular diameter, {\it Gaia} EDR3 parallax, surface gravity, binary mass function, and the geometric restriction that any star in a binary system has to be smaller than its Roche lobe (RL) radius.

\begin{figure}
\centering
\includegraphics[width=1\columnwidth]{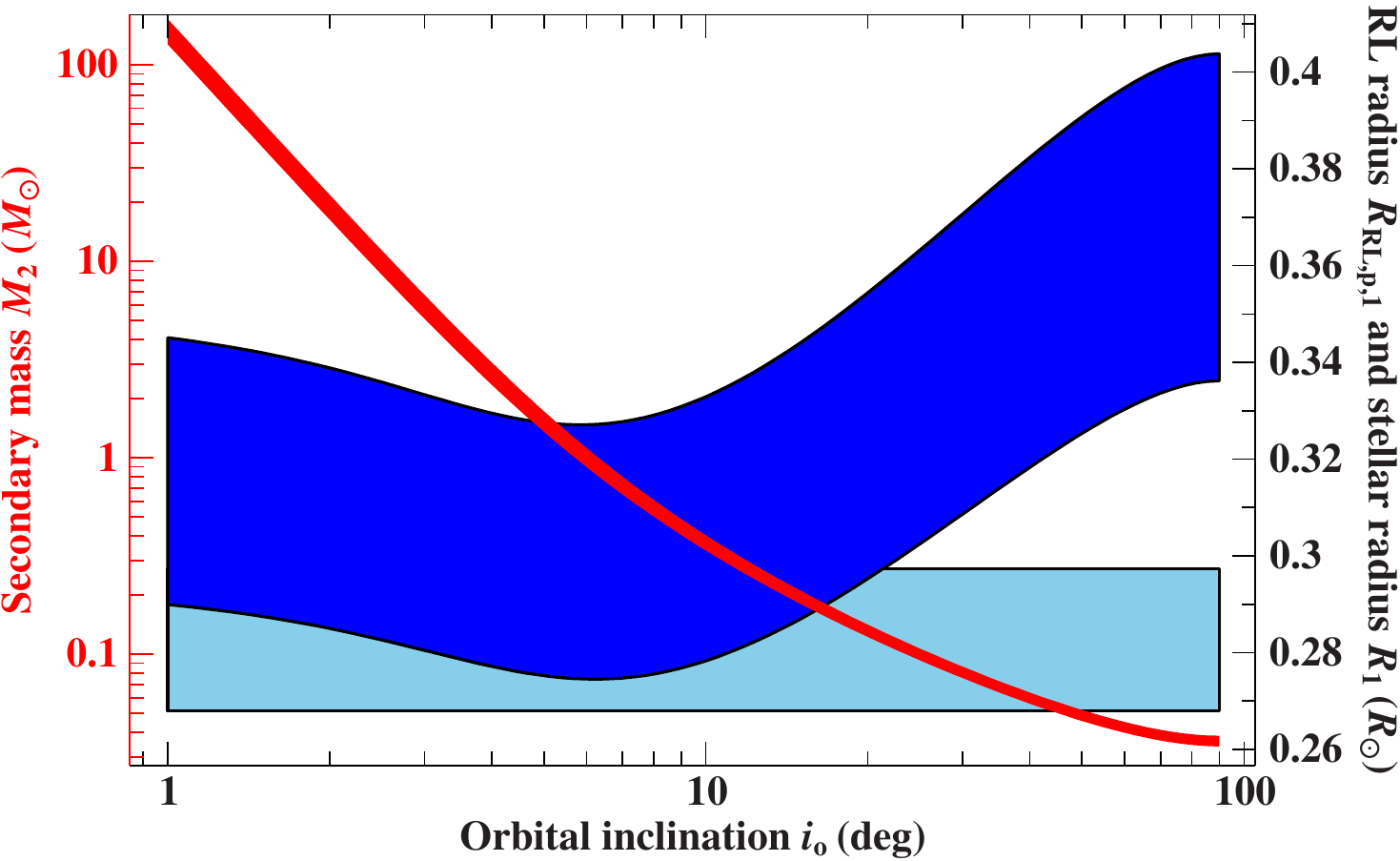}
\caption{\label{fig:mass_function}Mass and RL radius as a function of orbital inclination. Left ordinate (red): mass of the unseen companion, $M_2$, as a function of the orbital inclination, $i_\mathrm{o}$. A mass of $M_1=0.21\pm0.01\,M_\odot$ for the visible component is used to numerically solve the binary mass function given in Eq.~(\ref{eq:mass_function}) for $M_2$. Right ordinate (black-rimmed blue-shaded curves): comparison of the RL radius of the visible component at the pericenter passage based on Eq.~(\ref{eq:Roche_lobe_radius_periastron}), $R_{\mathrm{RL,p,}1}$ (dark blue), and the respective stellar radius, $R_1$ (light blue), following from $M_1$ and from the spectroscopically inferred surface gravity, $g=GM_1/R_1^2$ (see Table~\ref{table:atmospheric_parameters}; $G$ is the gravitational constant). Physical conditions are met when the light blue band is below the dark blue band, that is, when the star is smaller than its RL radius. The widths of all shaded regions cover all involved $1\sigma$ uncertainties. }
\end{figure}

To start with, we note that these four parameters are actually not independent of one another but are instead tightly coupled via the following two observational constraints: first, the value for the surface gravity,
\begin{equation}
g = GM_1/R_1^2 \,,
\label{eq:condition_g}
\end{equation}
as derived from spectroscopy ($G$ is the gravitational constant; see Table~\ref{table:atmospheric_parameters}), and second, the value for the mass function,
\begin{equation}
f = \frac{M_2 \sin^3(i_\mathrm{o})}{(1+M_1/M_2)^2} = (1-e^2)^{3/2}\frac{K_1^3 P}{2 \pi G} \,,
\label{eq:mass_function}
\end{equation}
as derived from the analysis of the radial-velocity curve (see Table~\ref{table:orbital_params}). Consequently, only two of the four quantities of interest have to be treated as free parameters in the statistical analysis. Because plausible prior information is available for $i_\mathrm{o}$ and $R_1$, we chose those two as independent variables and used Eqs.~(\ref{eq:condition_g}) and (\ref{eq:mass_function}) to (numerically) solve for the masses $M_1$ and $M_2$: $M_1 = M_1(g, R_1)$ and $M_2 = M_2(f, M_1(g, R_1), i_\mathrm{o}) = M_2(f, g, R_1, i_\mathrm{o})$ (see Fig.~\ref{fig:mass_function}). Assuming orbital inclinations to be distributed isotropically, the prior for $i_\mathrm{o}$ is simply
\begin{equation}
P(i_\mathrm{o}) \propto \sin(i_\mathrm{o}) \,.
\label{eq:prior_io}
\end{equation}
For the radius, $R_1$, we were able to construct a tailored prior from the measurements of the star's angular diameter ($\Theta = 3.82\times10^{-12}$ and $\sigma_\Theta = 0.06\times10^{-12}$; see Table~\ref{table:photometry_results}) and its {\it Gaia} EDR3 parallax ($\varpi = 0.117-Z$\,mas; $Z$ is the parallax bias or ``zero point''; $\sigma_\varpi=0.073$\,mas). \citetads{2020arXiv201203380L} advise a global parallax correction of $Z=-0.017$\,mas, while \citetads{2020arXiv201201742L} propose a more complex recipe that yields
$Z=-0.005$\,mas for this specific target. Because the difference between both values is significantly smaller than $\sigma_\varpi$, our results are insensitive to the choice of the correction scheme and we opted for $Z = -0.017$\,mas. Substituting the inverse distance, $1/d$, in the equation for the angular diameter, $\Theta = 2R_1/d$, with the parallax $\varpi = 1\mathrm{au}/d$ (au is the astronomical unit) and assuming $\Theta$ and $\varpi$ to be distributed like Gaussians yields the following integral for the prior probability of $R_1$:
\begin{eqnarray}
P(R_1) & \propto & \iint\limits_0^\infty\!\exp\left(-\frac{(\acute{\Theta}-\Theta)^2}{2\sigma_\Theta^2}-\frac{(\acute{\varpi}-\varpi)^2}{2\sigma_\varpi^2}\right)\delta\left(R_1-\frac{\acute{\Theta}\mathrm{au}}{2\acute{\varpi}}\right) \nonumber \\
& \times & P(\acute{\varpi})\mathrm{d}\acute{\Theta}\mathrm{d}\acute{\varpi} \,, \label{eq:prior_R1}
\end{eqnarray}
where $\delta$ is the Dirac delta function and $P(\varpi)$ is a prior for the parallax. Using a constant value for $P(\varpi)$ would lead to a long tail of unrealistically large radii because the parallax of J1604$+$1000 is only poorly constrained ($\sigma_\varpi/\varpi \approx 0.5$) and $R_1 \propto \varpi^{-1}$. One popular way of coping with this issue is to suppress small parallaxes by applying an exponentially decreasing distance prior as done, for example, by \citetads{2018AJ....156...58B}. However, as shown, for instance, by \citetads{2019A&A...627A.104D}, this approach is ``not well suited for analyzing individual halo stars when the parallax error is large.'' Therefore, in order to circumvent implausible parallaxes for our target, we had to pursue a different strategy. As outlined in Sect.~\ref{sect:intro}, large distances would render the star gravitationally unbound to the Milky Way and thus imply an extragalactic origin. Already for a single star, the probability of observing such a Galactic intruder would be extremely small, and it is even smaller for a binary system because its high space motion would require such violent acceleration scenarios that the binary system would rather be disrupted than ejected. Consequently, we introduced a prior for the parallax that favors Galactic trajectories that are bound to the Milky Way, that is, for which the difference between the current Galactic rest-frame velocity, $\varv_\mathrm{Grf}$, and the local Galactic escape velocity, $\varv_\mathrm{esc}$, is negative:
\begin{equation}
P(\varpi)\propto
\left\{\begin{array}{ll}
1\,, & \varv_\mathrm{Grf}-\varv_\mathrm{esc} \leq 0 \\
\exp\left(-(\varv_\mathrm{Grf}-\varv_\mathrm{esc})^2/2\sigma_\varv^2\right)\,, & \varv_\mathrm{Grf}-\varv_\mathrm{esc} > 0\,.
\end{array}
\right.
\label{eq:prior_parallax}
\end{equation}
Choosing this functional form and $\sigma_\varv = 10$\,km\,s$^{-1}$ ensures a smooth and gradual transition between bound and unbound orbits. The quantity $\varv_\mathrm{Grf}-\varv_\mathrm{esc}$ is a function of celestial coordinates, parallax, and proper motions, all of which are available in {\it Gaia} EDR3, as well as systemic radial velocity (see Table~\ref{table:orbital_params}) and the assumed Galactic mass distribution. For the latter, we employed Milky Way mass Model~I from \citetads{2013A&A...549A.137I}, which is consistent with Galactic mass estimates based on {\it Gaia} DR2 astrometry (see \citealt{2018A&A...620A..48I}). Uncertainties of and correlations between the other function arguments are accounted for in a Monte Carlo fashion when Eq.~(\ref{eq:prior_R1}) is numerically integrated to obtain the resulting prior distribution for $R_1$, which is shown as a solid gray line in Fig.~\ref{fig:corner_plots_Bayes}. The sharp drop at $R_1 \approx 0.46\,R_\odot$ is caused by Eq.~(\ref{eq:prior_parallax}) and corresponds to $\varpi = \Theta\mathrm{au}/(2R_1) = 3.82\times10^{-12}\times1\mathrm{au}/(2\times0.46\,R_\odot) = 0.18$\,mas, which coincides with the $1\sigma$ upper limit of the measured parallax. Consequently, the demand for bound trajectories requires the true parallax to be on the high side of what is currently allowed by {\it Gaia}.

\begin{table*}
\small
\centering
\renewcommand{\arraystretch}{1.25}
\setlength{\tabcolsep}{0.195cm}
\caption{\label{table:stellar_parameters}Parameters resulting from the Bayesian analysis as function of the RL filling factor, $x$.}
\begin{tabular}{c|ccccccccccccc}
\hline\hline
$x$ & $i_\mathrm{o}$ & $R_1$ & $M_1$ & $M_2$ & $M_1/M_2$ & $a$ & $R_{\mathrm{RL,p,}1}$ &  $R_{\mathrm{RL,p,}2}$ & $\varv_{\mathrm{grav},1}$ & $L_1/L_\odot$ & $d$ & $A_{\mathrm{elli},1}$ & $A_\mathrm{irra}$ \\
\cline{4-5} \cline{7-9} \cline{13-14}
& (deg) & $(R_\odot)$ & \multicolumn{2}{c}{$(M_\odot)$} & & \multicolumn{3}{c}{$(R_\odot)$} & (km\,s${}^{-1}$) & & (kpc) & \multicolumn{2}{c}{(mmag)} \\
\hline
1.0 & $90^{+\phantom{0}0}_{-43}$ & $0.43^{+0.03}_{-0.08}$ &    $0.41^{+0.10}_{-0.13}$ & $0.060^{+0.019}_{-0.019}$ &  $6.5^{+1.5}_{-2.0}$ & $0.89^{+0.13}_{-0.13}$ & $0.46^{+0.06}_{-0.08}$ & $0.20^{+0.03}_{-0.03}$ &    $0.66^{+0.11}_{-0.13}$ & $9.0^{+2.4}_{-2.8}$ & $5.0^{+0.4}_{-1.0}$ &                $13^{+6}_{-6}$ &           $2.3^{+0.6}_{-0.8}$ \\
0.9 & $90^{+\phantom{0}0}_{-37}$ &    $0.42^{+0.04}_{-0.08}$ &    $0.40^{+0.11}_{-0.13}$ & $0.060^{+0.019}_{-0.019}$ &  $6.5^{+1.5}_{-1.8}$ & $0.89^{+0.11}_{-0.11}$ & $0.46^{+0.06}_{-0.08}$ & $0.20^{+0.03}_{-0.03}$ &    $0.66^{+0.09}_{-0.13}$ & $8.6^{+2.4}_{-2.4}$ & $5.0^{+0.4}_{-1.0}$ &                $13^{+6}_{-5}$ &           $2.3^{+0.6}_{-0.7}$ \\
0.8 & $90^{+\phantom{0}0}_{-33}$ &    $0.38^{+0.07}_{-0.07}$ &    $0.34^{+0.14}_{-0.11}$ & $0.056^{+0.015}_{-0.015}$ &  $6.5^{+1.5}_{-1.3}$ & $0.89^{+0.11}_{-0.11}$ & $0.47^{+0.07}_{-0.09}$ & $0.20^{+0.02}_{-0.02}$ &    $0.64^{+0.13}_{-0.15}$ & $7.3^{+2.8}_{-2.1}$ & $4.4^{+0.9}_{-0.7}$ &                $13^{+5}_{-5}$ &           $2.4^{+0.6}_{-0.7}$ \\
0.7 & $90^{+\phantom{0}0}_{-29}$ &    $0.33^{+0.07}_{-0.06}$ &    $0.27^{+0.14}_{-0.09}$ & $0.048^{+0.013}_{-0.013}$ &  $6.5^{+1.3}_{-1.3}$ & $0.85^{+0.15}_{-0.11}$ & $0.44^{+0.09}_{-0.07}$ & $0.19^{+0.03}_{-0.02}$ &    $0.58^{+0.15}_{-0.13}$ & $5.6^{+2.8}_{-1.7}$ & $3.9^{+0.8}_{-0.7}$ &                $11^{+4}_{-4}$ &           $2.4^{+0.6}_{-0.6}$ \\
0.6 & $90^{+\phantom{0}0}_{-23}$ &    $0.29^{+0.07}_{-0.05}$ &    $0.22^{+0.13}_{-0.08}$ & $0.037^{+0.011}_{-0.009}$ &  $6.5^{+1.5}_{-1.5}$ & $0.85^{+0.13}_{-0.13}$ & $0.43^{+0.10}_{-0.07}$ & $0.19^{+0.03}_{-0.02}$ &    $0.52^{+0.15}_{-0.09}$ & $4.6^{+2.1}_{-1.7}$ & $3.4^{+0.8}_{-0.6}$ &           $\phantom{0}8^{+2}_{-2}$ &           $2.4^{+0.6}_{-0.6}$ \\
0.5 & $90^{+\phantom{0}0}_{-21}$ &    $0.27^{+0.07}_{-0.06}$ &    $0.19^{+0.12}_{-0.07}$ & $0.031^{+0.009}_{-0.009}$ &  $7.0^{+1.5}_{-1.5}$ & $0.89^{+0.22}_{-0.15}$ & $0.47^{+0.12}_{-0.09}$ & $0.20^{+0.03}_{-0.02}$ &    $0.50^{+0.13}_{-0.09}$ & $3.9^{+2.1}_{-1.4}$ & $3.2^{+0.8}_{-0.7}$ &           $\phantom{0}5^{+2}_{-2}$ &           $2.2^{+0.6}_{-0.6}$ \\
\hline
\end{tabular}
\tablefoot{The given numbers are the modes and the highest density intervals with probability 0.6827 (see \citeads{2018AJ....156...58B} for details on this measure of uncertainty) of the posterior PDFs computed via Monte Carlo integration. Corner plots and posterior PDFs for $i_\mathrm{o}$, $R_1$, $M_1$, and $M_2$ are shown in Fig.~\ref{fig:corner_plots_Bayes}. The quantity $a = 0.106 \pm 0.006\,R_\odot\,(1+M_1/M_2)/\sin(i_\mathrm{o})$ is the semimajor axis of the binary system, $R_{\mathrm{RL,p,}1}$ and $R_{\mathrm{RL,p,}2}$ are the RL radii of the visible and unseen component at pericenter passage, $\varv_{\mathrm{grav},1} = GM_1/(R_1c)$ is the gravitational redshift ($G$ is the gravitational constant, $c$ is the speed of light), $L_1/L_\odot = (R_1/R_\odot)^2 (T_\mathrm{eff}/T_\odot)^4$ is the luminosity, $d = 2R_1/\Theta$ is the distance, and $A_{\mathrm{elli},1}$ and $A_\mathrm{irra}$ are estimates for the expected $V$-band semi-amplitudes caused by ellipsoidal deformation and irradiance effects, respectively.}
\end{table*}
Independent of the above finding, there is another aspect that calls for even smaller stellar radii and thus turns out to be the main driver for high parallaxes: the simple geometric constraint that any star in a binary system has to be smaller than its RL radius. Owing to the compactness of the binary system, this requirement plays an important role because it renders low orbital inclinations and large stellar radii less likely. This assertion is illustrated in Fig.~\ref{fig:mass_function} for an exemplary mass of $M_1=0.21\pm0.01\,M_\odot$, but it is also valid for other masses. To be as general as possible, we accounted for the slight nominal eccentricity, $e$, of the binary orbit by focusing in the following on the orbital phase at which the RL is smallest, that is, on the periastron passage. The expression for the projected periastron distance is
\begin{eqnarray}
r_{\mathrm{p}} \sin(i_\mathrm{o}) & = & (1-e)\,(a_1 + a_2) \sin(i_\mathrm{o}) \nonumber \\
& = & (1-e)\,a_1 \sin(i_\mathrm{o})\,(1+a_2/a_1) \nonumber \\
& = & (1-e)\,a_1 \sin(i_\mathrm{o})\,(1+M_1/M_2)
,\end{eqnarray}
where $a_1$ and $a_2$ are the components' semimajor axes and
\begin{equation}
(1-e)\,a_1 \sin(i_\mathrm{o}) = (1-e)\,(1-e^2)^{1/2} \frac{K_1 P}{2\pi} = 0.102 \pm 0.008\,R_\odot
\end{equation}
is given from the analysis of the radial-velocity curve (see Table~\ref{table:orbital_params}). Using the shorthand notation $q \coloneqq M_1/M_2$ and an approximation formula for the relative RL radius \citepads{1983ApJ...268..368E},
\begin{equation}
r_\mathrm{RL}(q) = \frac{0.49\,q^{2/3}}{0.6\,q^{2/3}+\ln\left(1+q^{1/3}\right)} \,,
\end{equation}
the RL radius at periastron can be written as
\begin{eqnarray}
R_{\mathrm{RL,p},1}(f,g,R_1,i_\mathrm{o}) & = & r_\mathrm{RL}(q)\,r_{\mathrm{p}} = r_\mathrm{RL}(q) \frac{r_{\mathrm{p}} \sin(i_\mathrm{o})}{\sin(i_\mathrm{o})} \nonumber \\
& = & r_\mathrm{RL}(q) \frac{(1-e)\,a_1 \sin(i_\mathrm{o})\,(1+q)}{\sin(i_\mathrm{o})}
\label{eq:Roche_lobe_radius_periastron} \,.
\end{eqnarray}
The dependences on the mass function, $f$, the surface gravity, $g$, and the radius, $R_1$, enter via $q = M_1/M_2, M_1 = M_1(g, R_1)$, and $M_2 = M_2(f, g, R_1, i_\mathrm{o})$ (see above). Consequently, the geometric constraint that the stellar radius must not exceed the RL radius, $R_1 \leq x\,R_{\mathrm{RL,p,}1}$, yields
\begin{equation}
h(f, g, R_1, i_\mathrm{o}) \coloneqq \frac{R_1 \sin(i_\mathrm{o})}{r_\mathrm{RL}(q)(1+q)x} \leq (1-e)\,a_1 \sin(i_\mathrm{o}) \,.
\label{eq:condition_h}
\end{equation}
The factor $0 < x \le 1$ gives the fraction of the RL that the star is allowed to fill and is equal to 1 if\ only the current geometry of the binary system is considered. However, as discussed in Sect.~\ref{sect:visible_component}, such an approach would ignore the evolutionary history of the stellar components, which hints at $x<1$. To understand in detail how the Bayesian inference is affected by $x$, we did not marginalize our results over this parameter but explicitly explored different values for it in the following. This has the advantage that we do not bias the outcome by any prior assumption on $x$, which would be very vague at best.

The goal of the statistical analysis is to find those values of $i_\mathrm{o}$ and $R_1$ --~and thus also of $M_1$ and $M_2$~-- that have the highest probability of obeying the observational constraint given by Eq.~(\ref{eq:condition_h}). Hence, we wished to determine the probability of having a set of parameters $\{R_1,i_\mathrm{o}\}$ given the observation $h$, that is, we had to find an expression for the conditional probability $P(\{R_1,i_\mathrm{o}\}|h)$. Using Bayes' theorem, we can write
\begin{equation}
P(\{R_1,i_\mathrm{o}\}|h) = C\,P(h|\{R_1,i_\mathrm{o}\})\,P(R_1)\,P(i_\mathrm{o})
\label{eq:conditional_probability}
,\end{equation}
where $C$ is a normalization constant and $P(i_\mathrm{o})$ and $P(R_1)$ are the priors over the values of the two independent parameters that are given in Eqs.~(\ref{eq:prior_io}) and (\ref{eq:prior_R1}). The constraint from Eq.~(\ref{eq:condition_h}) is implemented via the following likelihood function:
\begin{equation}
P(h|\{R_1,i_\mathrm{o}\}) \propto
\left\{\begin{array}{ll}
1\,, & h \leq \bar{h} \\
\exp\left(-(h-\bar{h})^2/2\sigma_h^2\right)\,, & h > \bar{h}\,.
\end{array}
\right.
\end{equation}
The quantities $\bar{h} = 0.102\,R_\odot$ and $\sigma_h = 0.008\,R_\odot$ are derived from the observed radial-velocity curve (see Table~\ref{table:orbital_params}). With all expressions in Eq.~(\ref{eq:conditional_probability}) defined, it is possible to compute the posterior probability distribution functions (PDFs) for the two independent variables by marginalizing out the other one. To this end, we applied a Monte Carlo procedure for the numerical integration, which allowed us to propagate the uncertainties in the surface gravity, $g$, and in the mass function, $f$. Posterior PDFs for various derived quantities, such as $M_1$ and $M_2$, were also computed.

The result of this exercise for RL filling factors $0.5 \le x \le 1$ is summarized in Table~\ref{table:stellar_parameters} and illustrated in Fig.~\ref{fig:corner_plots_Bayes}. Scenarios with $x<0.5$ are not discussed here because they predict stellar radii that are highly unlikely given the prior information on $R_1$. Interestingly, the mass ratio, the semimajor axis, and the RL radii of the two components are almost insensitive to the assumed value for $x$. Moreover, independent of $x$, the unseen companion is most likely not massive enough to ignite hydrogen burning in its core and therefore qualifies as a candidate BD. In contrast, the inferred parameters of the visible component are sensitive to the precise value of $x$. Here we first consider the case where the star is allowed to fill its entire current RL at periastron, that is, $x=1$. As demonstrated in Fig.~\ref{fig:corner_plots_Bayes}, the prior and posterior PDFs of $i_\mathrm{o}$ and $R_1$ are then almost identical, which shows that the geometric constraint of Eq.~(\ref{eq:condition_h}) barely affects the outcome in this case. This is expected when looking at Fig.~\ref{fig:mass_function}: Requiring the RL radius to exceed the stellar radius only renders small orbital inclinations and large stellar radii a little less likely, the latter of which are in any case already unlikely by virtue of Eq.~(\ref{eq:prior_parallax}). Consequently, $R_1$ and thus also $M_1$ are essentially prescribed by the prior on $R_1$ (see Eq.~(\ref{eq:prior_R1})). We now turn to cases with $x<1$. As demonstrated in Fig.~\ref{fig:corner_plots_Bayes}, this requirement puts increasingly strong constraints on the four parameters of interest. Generally speaking, the smaller $x $ is, the smaller $M_1$, $M_2$, and $R_1$ are and the larger $i_\mathrm{o}$ is, which is not surprising because it simply reflects the fact that a smaller allowed volume necessitates a smaller object. In principle, the light curve could be exploited to constrain $x,$ and thus the precise parameters of the visible component, solely from observations. To demonstrate this, we calculated the expected $V$-band semi-amplitudes caused by the ellipsoidal deformation of the visible component, $A_{\mathrm{elli},1}$, and by irradiance effects, $A_\mathrm{irra}$, following Eqs.~(3) and (6) in \citetads{1993ApJ...419..344M}. A gravity-darkening coefficient of $\tau_1 =  0.51$ (see Eq.~(10) in \citeads{1985ApJ...295..143M}) and a limb-darkening coefficient of $u_1 = 0.34$ \citepads{2011A&A...529A..75C} were used to compute $A_{\mathrm{elli},1}$. To calculate a rough upper limit for $A_\mathrm{irra}$, the unknown stellar radius and effective temperature of the unseen companion were approximated by their highest plausible values, that is, by $R_{\mathrm{RL,p,}2}$ and by assuming a very hot BD of $3000$\,K \citepads{2003A&A...402..701B}, respectively. Furthermore, blackbody behavior was assumed to compute the magnitude
difference between the two components, which enters the formula for $A_\mathrm{irra}$. The resulting numbers, which are listed in Table~\ref{table:stellar_parameters}, show that ellipsoidal deformation is the dominating effect and that its strength decreases from about 13 mmag to 5\,mmag when going from $x=1$ to $x=0.5$. While the currently available light curve does not allow us to resolve semi-amplitudes below $\sim 110$\,mmag (see Sect.~\ref{sect:lightcurve}), future follow-up observations might be good enough to measure those subtle variations with the required precision. For the time being, we have to rely on aspects from stellar evolution theory to narrow down $x$ and thus argue that J1604$+$1000 hosts a proto-He WD (see Sect.~\ref{sect:visible_component}).
\subsection{\label{sect:kinematic_analysis}Kinematic analysis}
\begin{figure}
\centering
\includegraphics[width=1\columnwidth]{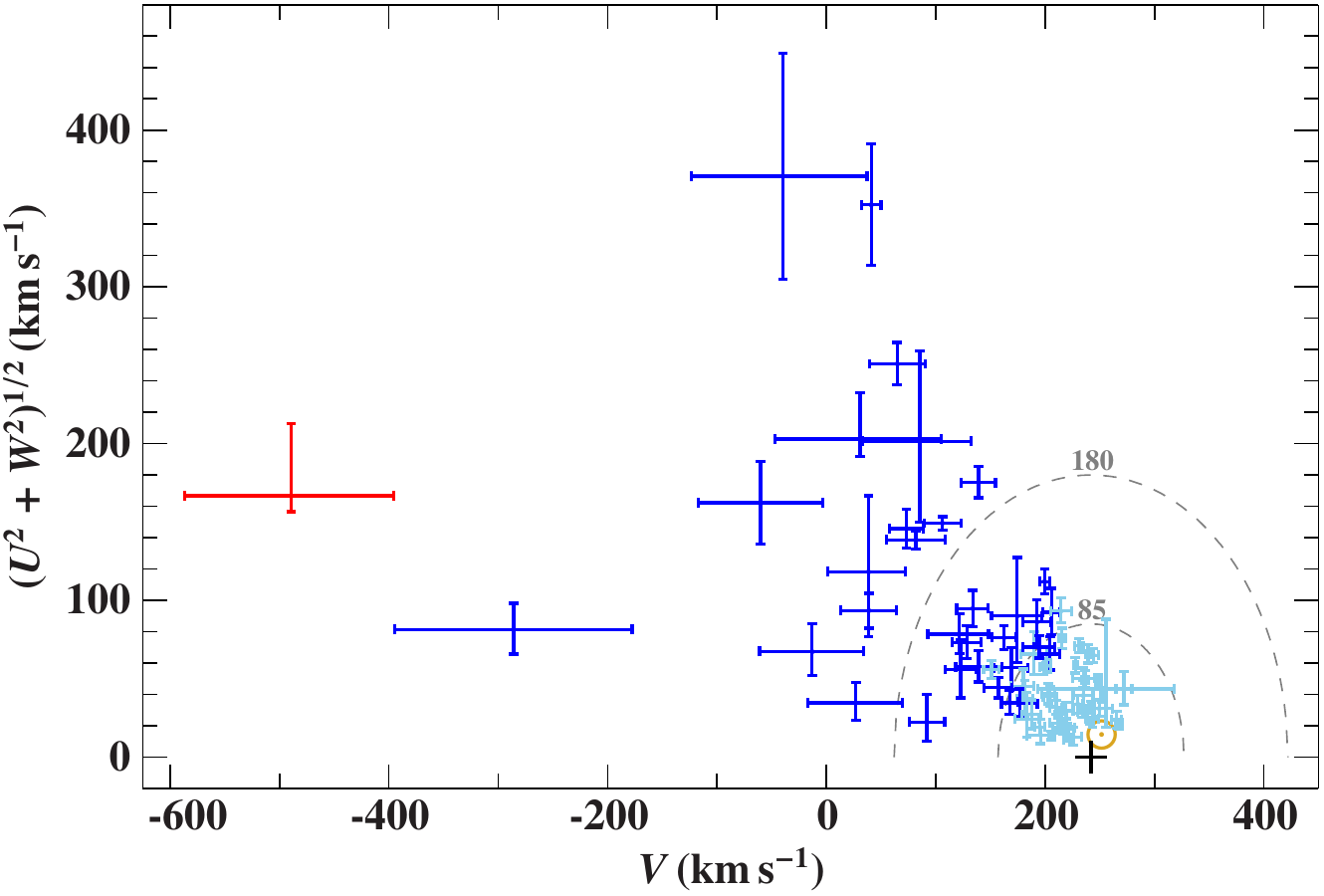}
\caption{\label{fig:toomre}Position of the star in the Toomre diagram for $x=0.8$. The quantity $V$ is the velocity component in the direction of Galactic rotation, $U$ is in the direction towards the Galactic center, and $W$ is perpendicular to the Galactic plane. The star, the Sun, and the local standard of rest (LSR) are marked by a red cross with $1\sigma$ error bars, a yellow circled dot ($\odot$), and a black plus sign ($+$), respectively. The dashed gray circles centered around the LSR represent boundaries for thin (radius of $85$\,km\,s${}^{-1}$) and thick (radius of $180$\,km\,s${}^{-1}$) disks following \citetads{2004AN....325....3F}. A reference sample of 75 binaries that host extremely low-mass He WDs is shown as blue data points, with light blue marking objects that orbit in the Galactic disk \citepads{2020ApJ...889...49B}.}
\end{figure}
The unusually large negative systemic radial velocity (see Table~\ref{table:orbital_params}) of this binary system calls for a closer inspection of its kinematic properties. Using the spectroscopic distance and the correction for gravitational redshift from Table~\ref{table:stellar_parameters} (for an exemplary case of $x=0.8$), as well as proper motions from {\it Gaia} EDR3, which seem to be reliable because the renormalized unit weight error (RUWE; see \citeads{RUWE}) indicates a well-behaved astrometric solution ($\text{RUWE}=0.99$), allows the current position and velocity vector of the system to be computed. Based on the object's location in the Toomre diagram (Fig.~\ref{fig:toomre}), we conclude that it is on a highly retrograde, halo-like orbit that is more extreme than that of similar binary systems. The fact that the respective difference between the current Galactic rest-frame velocity and the local Galactic escape velocity, $\varv_{\mathrm{Grf}}-\varv_{\mathrm{esc}}=-150^{+120}_{-\phantom{0}80}$\,km\,s${}^{-1}$, is well below zero illustrates that the condition $x=0.8$ is already more than enough to make the system gravitationally bound to the Milky Way, even without imposing this condition via Eq.~(\ref{eq:prior_parallax}).
\section{Discussion}
\subsection{The nature of the unseen component}
The companion of the visible B-type star does not exhibit signatures in the optical spectra. Despite this lack of a direct hint, the available constraints clearly indicate that the object is of very low mass (see Table~\ref{table:stellar_parameters}), most probably below the metallicity-dependent hydrogen burning limit of $\sim$0.07--0.094$\,M_\odot$ (see, e.g., \citeads{2014AJ....147...94D} and references therein). Consequently, it is a BD candidate. This is primarily a result of the system's low mass function (see Table~\ref{table:orbital_params}) and the prior distribution for $i_\mathrm{o}$, which, according to Eq.~(\ref{eq:prior_io}), favors high inclinations and consequently low companion masses (see Fig.~\ref{fig:mass_function}). For the $M_2$ values listed in Table~\ref{table:stellar_parameters}, models by \citetads{2003A&A...402..701B} predict BD radii of 0.12--0.17\,$R_\odot$ for an age of 0.1\,Gyr and 0.09--0.10\,$R_\odot$ for an age of 10\,Gyr, all of which are smaller than the corresponding RL radius $R_{\mathrm{RL,p,}2}$, which is in the range 0.17--0.23$\,R_\odot$ (see Table~\ref{table:stellar_parameters}). So even without invoking any restrictions on the dimension of the companion, a consistent picture is obtained when assuming a BD nature for the unseen component.

In principle, the SED could be exploited to further investigate the nature of the low-mass companion, substellar or not, owing to its expected low temperature ($\sim$1500--3000\,K) that makes the object shine in the infrared. However, even when assuming the lowest $R_1$ value from Table~\ref{table:stellar_parameters}, $R_1 = 0.27^{+0.07}_{-0.06}\,R_\odot$ for $x=0.5$, the maximum possible relative radiation area of the unseen component is $(R_{\mathrm{RL,p,}2}/R_1)^2 = 0.38^{+0.38}_{-0.18}$, which is almost two orders of magnitude lower than what is obtained from modeling the observed infrared excess with a blackbody component ($17^{+16}_{-\phantom{0}5}$; see Table~\ref{table:photometry_results}). We therefore conclude that the observed infrared excess is much too strong to be explained by the low-mass component, which also renders the prospects of obtaining insights about the companion from SED modeling dim. Instead, optical light curves with higher cadence and precision than what is currently available are eventually necessary in order to search for potential signatures of the companion, such as ellipsoidal modulation or an irradiance effect, which are estimated to be in the millimagnitude regime (see Table~\ref{table:stellar_parameters}). The unexpected presence of an infrared excess may be explained as a relic of a previous mass transfer episode (see Sect.~\ref{sect:thin_hydrogen_envelope}).
\subsection{\label{sect:visible_component}The nature of the visible component}
\begin{figure}
\centering
\includegraphics[width=0.99\columnwidth]{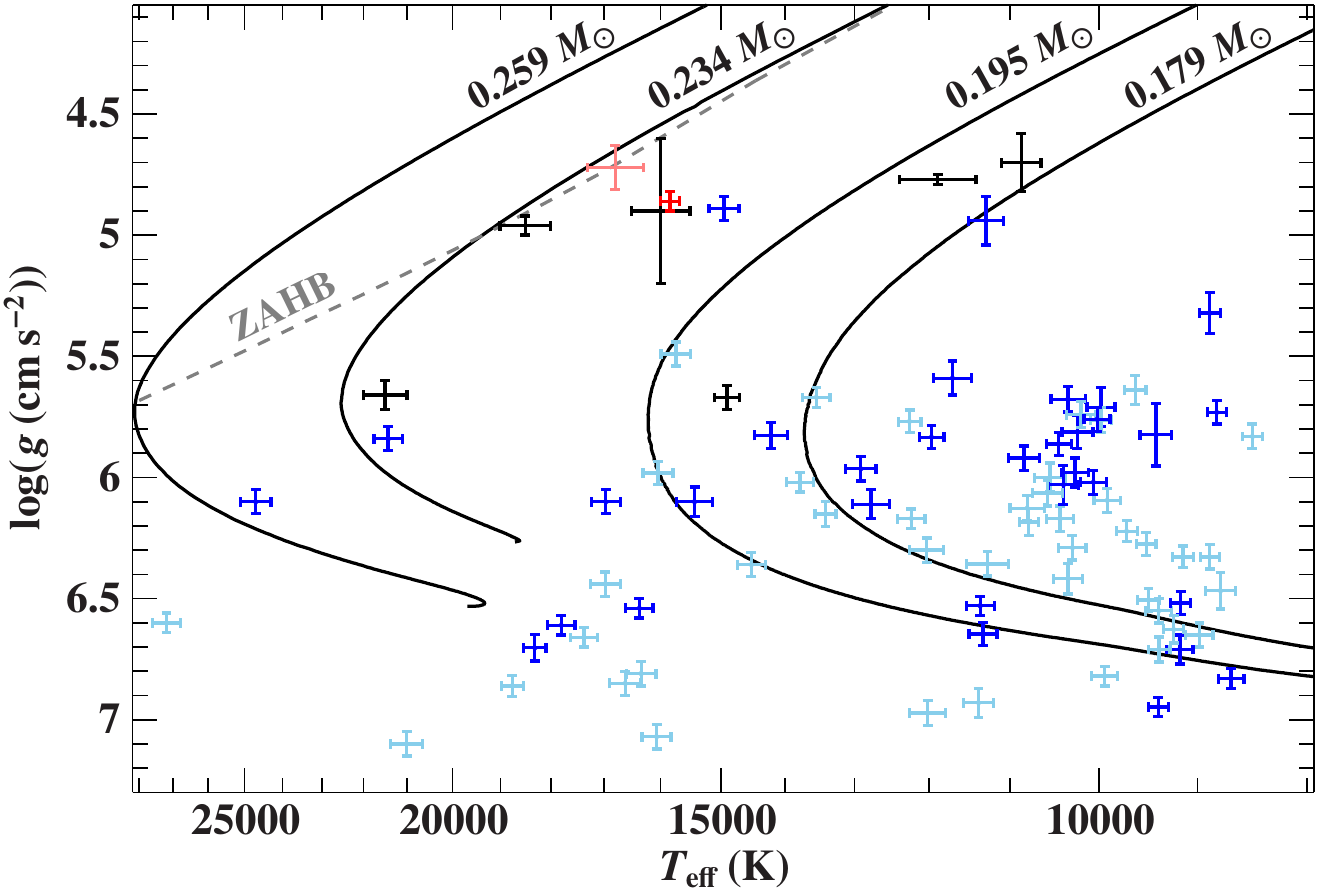}
\caption{\label{fig:evolution_tracks}Preliminary and revised position of J1604$+$1000 (light red and dark red $1\sigma$ error bars) in the Kiel diagram. The black lines are evolutionary tracks for stripped helium cores from \citetads{1998A&A...339..123D} labeled with their respective masses (stars evolve from top to bottom). For reference, the locus of the zero-age horizontal branch (ZAHB) for $[\mathrm{Fe}/\mathrm{H}]=-1.48$ \citepads{1993ApJ...419..596D} is shown as a dashed gray line. The meaning of the (light) blue data points is the same as in Fig.~\ref{fig:toomre}. The six proto-He WDs that are mentioned in the text -- PSR\,J1816$+$4510, EVR-CB-001, GALEX\,1717$+$6757, HD\,188112, EL\,CVn, and WASP\,0247$-$25\,B -- are presented in black.}
\end{figure}
The analysis of the follow-up spectra yielded improved atmospheric parameters with respect to the preliminary analysis outlined in Sect.~\ref{sect:intro}. The effective temperature turned out to be lower and the surface gravity higher. Figure~\ref{fig:evolution_tracks} shows the preliminary and revised position of J1604$+$1000 in the Kiel diagram, demonstrating that the new parameters are no longer consistent with those of a BHB star because they place the star well below the zero-age horizontal branch (ZAHB). Instead, the comparison with evolutionary tracks for stripped helium cores suggests that the star is a proto-He WD of $0.21\pm0.01\,M_\odot$, which, according to Table~\ref{table:stellar_parameters}, would correspond to $0.5 \lesssim x \lesssim 0.6$. However, as outlined in Sect.~\ref{sect:thin_hydrogen_envelope}, the application of those tracks has one major caveat for this particular object, which is why we refrain from using them to determine the stellar parameters.

The evolutionary histories of BHB stars and He WDs are very similar. The horizontal branch (HB) is a sequence of core-helium burning stars that formed after the ignition of helium burning at the tip of the red giant branch (RGB). The morphology of the HB is determined by the mass of the hydrogen envelope: The lower it is, the bluer the HB star appears. Hence, in order to populate the BHB, a considerable fraction of the envelope mass must have been lost during the transition from the RGB to the HB. However, it is conceivable that mass loss occurs even before the progenitor star has reached the tip of the RGB and ignited helium burning in the core. The outcomes of such an early mass loss are stripped stars that cool down and become low-mass He WDs. If they are less massive than 0.3\,$M_\odot$, they are called extremely low-mass (ELM) WDs. The high mass loss that is necessary to strip off the envelope and eventually form He WDs is most likely due to binary mass transfer (\citeads{1998A&A...339..123D}, \citeads{2013A&A...557A..19A}, \citeads{2016A&A...595A..35I}). In fact, most ELM WDs are found in short-period double-degenerate systems (see \citeads{2020ApJ...889...49B} and references therein).

Although the surface gravity of J1604$+$1000 is currently not high enough to classify it as a WD, its low mass and its status as short-period binary system make it a prime candidate for being the progenitor of a low-mass He WD. This assumption is strengthened by noting that the atmospheric parameters of our  star are almost identical to those of the low-mass proto-He WD companion of the millisecond pulsar PSR\,J1816$+$4510 ($T_\mathrm{eff}=16\,000\pm500$\,K, $\log(g)=4.9\pm0.3$; \citeads{2013ApJ...765..158K}). Another He WD progenitor with similarly low surface gravity but slightly higher temperature is EVR-CB-001 \citepads{2019ApJ...883...51R}, which is a binary system composed of a pre-ELM WD ($T_\mathrm{eff}=18\,500\pm500$\,K, $\log(g)=4.96\pm0.04$) and an unseen low-mass He WD companion. Finding such unevolved proto-He WDs offers the opportunity to study a key intermediate stage in their evolution \citepads{2019ApJ...883...51R}.

Assuming the visible component to be a proto-He WD now provides an argument for why to investigate RL filling factors $x<1$ (see Sect.~\ref{sect:Bayesian_inference}). Proto-He WDs generally contract with time (see, e.g., evolutionary tracks by \citeads{1998A&A...339..123D}, \citeads{2013A&A...557A..19A}, \citeads{2014A&A...571A..45I}, and \citeads{2016A&A...595A..35I}), and it could well be that the stripping event that formed the proto-He WD did not occur recently but rather several million years ago, giving the stripped star time to shrink by some fraction in the meantime. Because the typical timescales for changes in the binary orbit by the emission of gravitational waves or by orbital interactions with a potential third body, such as a shell or disk, are much longer, we can assume that the orbital configuration has remained more or less unaltered since the stripping event. We may thus not only require that the current stellar radius not exceed the RL radius but also that a somewhat larger object had to have fit into the current orbital geometry in the past. This is equivalent to saying that the current shrunken stellar radius has to be smaller than an unknown fraction, $x$, of the RL radius.
\subsection{Standard formation scenarios for He WDs}
Typically, (proto-)He WDs are found in one of the three following systems: double-degenerate systems (see, e.g., \citeads{2020ApJ...889...49B} and references therein), where both components are WDs; in millisecond pulsar systems (see, e.g., \citeads{2014A&A...571L...3I} and references therein), where the (proto-)WD orbits a fast spinning neutron star; and in EL\,CVn binaries \citepads{2014MNRAS.437.1681M}, where the proto-He WD is the under-luminous B-type star in an eclipsing binary systems with A- or F-type main sequence primaries (see, e.g., \citeads{2017MNRAS.467.1874C}, \citeads{2020ApJ...888...49W} and references therein). While millisecond pulsar systems and EL\,CVn binaries result from mass transfer onto the companion through stable RL overflow, the formation of double-degenerate systems depends on the mass ratio of the accreting WD and the donor star, which usually is an RGB star of $\sim$1--2$\,M_\odot$. If the WD accretor is massive enough to yield a mass ratio that is sufficiently close to unity, mass will be transferred via stable RL overflow. In contrast, if the mass ratio is significantly different from unity, mass transfer will not be stable and a common-envelope (CE) phase will occur (see, e.g., \citeads{2016A&A...595A..35I} and references therein for more details).

Given that CE evolution is one of the most complex and uncertain aspects in stellar evolution theory, it is not surprising that most theoretical studies about the evolution of low-mass He WDs focus on the RL overflow channel, either by explicitly following the evolution of a binary (e.g., \citeads{2013A&A...557A..19A}, \citeads{2014A&A...571A..45I}, \citeads{2016A&A...595A..35I}) or by artificially removing the envelope during the calculation of single-star evolutionary models (see, e.g., \citeads{1998A&A...339..123D}). As long as the initial model for the stripped helium core is correct, the latter approach is justified by noting that the further evolution of the model is independent of the details of the previous mass loss episode \citepads{1998A&A...339..123D}. Consequently, the two channels, RL overflow and CE ejection, may yield very similar evolutionary tracks in the Kiel diagram, provided that the post-stripping structure is indeed similar. A similar post-stripping structure is, for example, assumed in the binary population synthesis study by \citetads{2019ApJ...871..148L}, who discuss the characteristics of double degenerates resulting from the two formation channels.

However, a key aspect for the further evolution of a recently stripped proto-He WD is $M_\mathrm{H}$, the mass of its hydrogen envelope (see, e.g., \citeads{2000MNRAS.316...84S}). Calculations with RL overflow --~and thus almost all existing evolutionary tracks~-- robustly predict relatively thick hydrogen envelopes ($M_\mathrm{H}\sim10^{-4}$--$10^{-2}\,M_\odot$; see, e.g., \citeads{1998A&A...339..123D}, \citeads{2013A&A...557A..19A}, \citeads{2016A&A...595A..35I}). For CE evolution, the situation is more complex because the details of this dynamical process are far from being understood, which is why $M_\mathrm{H}$ remains an unknown parameter in that case. According to \citetads{2018A&A...614A..49C}, the formation of low-mass He WDs with thin hydrogen envelopes ($M_\mathrm{H}\sim10^{-6}\,M_\odot$) can then not be discarded, although strong observational evidence for that is currently missing. The proposed smoking gun for this scenario is the discovery of ELM WDs with effective temperatures below 7000\,K \citepads{2018A&A...614A..49C}. In the following, we suggest that J1604$+$1000 has formed with a thin hydrogen envelope, probably providing the as yet missing evidence for this hypothesis.
\subsection{\label{sect:thin_hydrogen_envelope}The need for a thin hydrogen envelope}
Based on the discussions in the previous sections, we propose the following formation scenario for J1604$+$1000, which has to be fine-tuned by tailored binary evolution calculations that are beyond the scope of this discovery paper. The progenitor system is composed of a low-mass ($\sim$1--2$\,M_\odot$) main sequence star plus a probably substellar companion of much lower mass ($\lesssim0.08\,M_\odot$). Triggered by stellar evolution, the primary expands and, at some point, fills its RL, which leads to the formation of a CE owing to the extreme mass ratio. Friction forces cause the system to spiral in and to finally eject the CE, leaving behind a close binary system that is surrounded by the expanding stripped envelope. The signature of a cool extended source observed as infrared excess in the analysis of the SED (see Sect.~\ref{sect:photometry}) may then be caused by this circumstellar material. A similar scenario has been suggested for the formation of WDs with BD companions (see \citeads{2017MNRAS.471..976P} and references therein).

The star's position in the Kiel diagram combined with evolutionary tracks with thick hydrogen envelopes (\citeads{1998A&A...339..123D}, \citeads{2013A&A...557A..19A}, \citeads{2014A&A...571L...3I}) hints at a mass for the stripped helium core of about $0.21\pm0.01\,M_\odot$ (see Fig.~\ref{fig:evolution_tracks}). However, those tracks also predict $\log(g)\sim3$ right after stripping, which would imply that the radius of the exposed core has decreased by a factor of $\sim10^{(4.86-3)/2}=8.5$ since the stripping event. Because a $0.21\pm0.01\,M_\odot$ star with $\log(g)=4.86\pm0.04$ is already close to filling its RL (see Fig.~\ref{fig:mass_function}), it is impossible to make such an object considerably larger in the past, which rules out this scenario. As demonstrated by \citetads{2018A&A...614A..49C}, a reduction in the thickness of the hydrogen envelope causes the envelope to become denser, which leads to a smaller stellar radius and thus to a larger surface gravity. Consequently, evolutionary tracks with thin hydrogen envelopes may resolve this contradiction and provide a spectroscopic mass estimate in a fully consistent picture.
\subsection{Abundance studies of (proto-)He WDs}
Regardless of the precise mass of J1604$+$1000, it is one of the very rare cases that have allowed for a comprehensive abundance analysis of a proto-He WD. Little information on the chemical composition of other (proto-)He WDs is currently available, which remains a crucial barrier to the advancement of our understanding of these objects (see, e.g., \citeads{2016A&A...595A..35I}). For the companion of the abovementioned millisecond pulsar PSR\,J1816$+$4510, \citetads{2013ApJ...765..158K} determined super-solar abundances for He, Na, Mg, Si, Ca, and Fe, which, apart from Fe, is completely contrary to what we find here. This is remarkable given that the atmospheric parameters of the two visible components are almost identical. For the ELM WD SDSS\,J074511.56$+$194926.5 ($T_\mathrm{eff}=8380\pm120$\,K, $\log(g)=6.21\pm0.07$), \citetads{2014ApJ...781..104G} measured solar abundances for Mg, Ca, Ti, Cr, and Fe. The most comprehensive abundance studies available are based on ultraviolet spectra of GALEX\,J1717$+$6757 (\citeads{2011ApJ...737L..16V}, \citeads{2014MNRAS.444.1674H}), a pre-ELM WD of similar temperature ($14\,900\pm200$\,K) but higher surface gravity ($5.67\pm0.05$\,dex) than our star, and the sdB-type pre-ELM WD HD\,188112 ($T_\mathrm{eff}=21\,500\pm500$\,K, $\log(g)=5.66\pm0.06$; \citeads{2003A&A...411L.477H}, \citeads{2016A&A...585A.115L}). \citetads{2014MNRAS.444.1674H} derived abundances of C, Al, Si, P, S, Ca, Ti, Cr, and Fe as well as upper limits for N, O, Mg, Sc, and Ni, while \citetads{2016A&A...585A.115L} determined abundances of Mg, Al, Si, P, S, Ca, Ti, Cr, Mn, Fe, Ni, Zn, Ga, Sn, and Pb as well as upper limits for C, N, and O. The resulting abundance patterns are thought to be produced by an interplay of atomic diffusion and rotational mixing (\citeads{2014MNRAS.444.1674H}, \citeads{2016A&A...585A.115L}, \citeads{2016A&A...595A..35I}). Also, based on ultraviolet spectroscopy, \citetads{2020AJ....159....4W} recently found evidence for atomic diffusion in the atmosphere of the pre-ELM WD EL\,CVn ($T_\mathrm{eff}=11\,890\pm490$\,K, $\log(g)=4.77\pm0.02$). \citetads{2017ApJ...847..130I} and \citetads{2018PhDT.......103H} presented an abundance analysis of the EL CVn-type star WASP\,0247$-$25\,B ($T_\mathrm{eff}=10\,870\pm230$\,K, $\log(g)=4.70\pm0.12$), which turned out to be somewhat rich in He and deficient in O, Mg, Si, Ca, Ti, and Fe, though these deficiencies are less pronounced than those for J1604$+$1000. Finally, based on a visual inspection of spectra of the ELM survey \citepads{2020ApJ...889...49B}, \citetads{2014MNRAS.444.1674H} concluded that all ELM WDs show Ca in their optical spectra if their surface gravity is lower than $\log(g)=5.9$.
\section{Summary and outlook}
Based on a comprehensive investigation that utilizes multi-epoch data from astrometry, photometry, and optical spectroscopy, we infer that J1604$+$1000 is a short-period ($P \approx 3.4$\,h) single-lined spectroscopic binary system that contains a visible B-type star ($M_1\lesssim0.52\,M_\odot$) and an unseen low-mass companion ($M_2\lesssim0.08\,M_\odot$) that is most likely a BD. The masses of the two components were constrained via Bayesian inference from the measured angular diameter, {\it Gaia} EDR3 parallax, surface gravity, binary mass function, and the condition that the stellar radius has to be smaller than its respective RL radius. The combination of effective temperature, $T_{\mathrm{eff}}=15\,840\pm160$\,K, surface gravity, $\log(g)=4.86\pm0.04$, and stellar mass suggests that the visible component is a proto-He WD. It exhibits a peculiar abundance pattern that is indicative of ongoing atomic diffusion processes. In terms of number fractions, He and Mg are below solar by more than 1\,dex and Ca by 0.7\,dex, while Fe is above solar by about 0.8\,dex. Although other chemical species do not exhibit spectral lines in the available optical spectra, upper abundance limits for C, N, O, Ne, Al, Si, S, and Ar can be derived; these elements are also significantly under-abundant (from 0.5\,dex for N to 2.5 dex for Si) with the exception of Ar, for which the limit is close to solar. Besides providing an estimate for the angular diameter, $\Theta = 2R_1/d = (3.82\pm0.06)\times10^{-12}$ ($R_1$ is the radius of the visible component and $d$ the distance to the binary system), the investigation of the SED also reveals an infrared excess that can be empirically modeled by a blackbody component with temperature $T_{\mathrm{bb}} = 2300^{+400}_{-600}$\,K and an effective radiation area that is $17^{+16}_{-\phantom{0}5}$ times the projected surface area of the visible star.

All observational constraints can be consistently explained by proposing that the system is a post-CE binary that probably emerged relatively recently from the mass transfer phase. Its extremely high Galactic rest-frame velocity indicates that it belongs to a very old (halo) population. Therefore, the BD is probably compact, which favored its prospects to survive its engulfment in the envelope of the red giant progenitor. Tighter constraints on the stellar parameters of the two components can be expected from more precise and accurate parallax measurements from future {\it Gaia} data releases as well as from optical light curves that are at least ten times more precise than the currently available CSS $V$-band light curve; this would allow the effect of ellipsoidal deformation to be measured. Moreover, owing to the compactness of the binary and the inferred constraints on the orbital inclination, the system is very likely eclipsing, with primary eclipse depths of roughly 10\%.

The reported discovery is particularly interesting for two reasons. First, the derived chemical abundance pattern of the proto-He WD shows heavy signatures of ongoing atomic diffusion processes and may thus help us to better understand the details of those complex processes, especially because quantitative abundance studies of proto-He WDs have been very rare so far. High-resolution ultraviolet spectroscopy would allow for abundance determinations of many more chemical species than what is currently possible based on optical spectra. Second, standard evolutionary tracks for stripped helium cores predict post-stripping stellar radii that are too large to be consistent with the current orbital configuration of J1604$+$1000, thus making the object a promising test bed for theory. Motivated by the RL overflow scenario, all of those standard evolutionary tracks are computed for thick hydrogen envelopes. However, this approach may not be valid for systems that underwent a CE phase, as we propose for J1604$+$1000. Following \citetads{2018A&A...614A..49C}, we argue that models with thin hydrogen envelopes would probably resolve the abovementioned inconsistency because they predict smaller stellar radii and thus larger surface gravities. Consequently, the thickness of the hydrogen envelope should be considered as an additional parameter in the spectroscopic mass and age determinations of (proto-)He WDs in post-CE systems.
\begin{acknowledgements}
We thank the anonymous referee for a very careful and constructive report. A.I.\ and U.H.\ acknowledge funding by the Deutsche For\-schungs\-gemeinschaft (DFG) through grants IR190/1-1 and HE1356/71-1. S.G.\ was supported by the Heisenberg program of the DFG through grants GE 2506/8-1 and GE 2506/9-1. This research was supported in part by the National Science Foundation under Grant No.\ NSF PHY-1748958. We thank John E.\ Davis for the development of the {\sc slxfig} module used to prepare the figures in this paper. Based on observations made with ESO Telescopes at the La Silla Paranal Observatory under programme ID 0102.D-0092(A). The William Herschel Telescope is operated on the island of La Palma by the Isaac Newton Group of Telescopes in the Spanish Observatorio del Roque de los Muchachos of the Instituto de Astrofísica de Canarias. The ISIS spectroscopy was obtained as part of W17AN011. Some of the data presented herein were obtained at the W.M.\ Keck Observatory, which is operated as a scientific partnership among the California Institute of Technology, the University of California and the National Aeronautics and Space Administration. The Observatory was made possible by the generous financial support of the W.M.\ Keck Foundation. The authors wish to recognize and acknowledge the very significant cultural role and reverence that the summit of Mauna Kea has always had within the indigenous Hawaiian community. We are most fortunate to have the opportunity to conduct observations from this mountain. This work has made use of data from the European Space Agency (ESA) mission {\it Gaia} (\url{https://www.cosmos.esa.int/gaia}), processed by the {\it Gaia} Data Processing and Analysis Consortium (DPAC, \url{https://www.cosmos.esa.int/web/gaia/dpac/consortium}). Funding for the DPAC has been provided by national institutions, in particular the institutions participating in the {\it Gaia} Multilateral Agreement. Funding for SDSS-III has been provided by the Alfred P.\ Sloan Foundation, the Participating Institutions, the National Science Foundation, and the U.S.\ Department of Energy Office of Science. The SDSS-III web site is \url{http://www.sdss3.org/}. SDSS-III is managed by the Astrophysical Research Consortium for the Participating Institutions of the SDSS-III Collaboration including the University of Arizona, the Brazilian Participation Group, Brookhaven National Laboratory, Carnegie Mellon University, University of Florida, the French Participation Group, the German Participation Group, Harvard University, the Instituto de Astrofisica de Canarias, the Michigan State/Notre Dame/JINA Participation Group, Johns Hopkins University, Lawrence Berkeley National Laboratory, Max Planck Institute for Astrophysics, Max Planck Institute for Extraterrestrial Physics, New Mexico State University, New York University, Ohio State University, Pennsylvania State University, University of Portsmouth, Princeton University, the Spanish Participation Group, University of Tokyo, University of Utah, Vanderbilt University, University of Virginia, University of Washington, and Yale University. The Pan-STARRS1 Surveys (PS1) have been made possible through contributions of the Institute for Astronomy, the University of Hawaii, the Pan-STARRS Project Office, the Max-Planck Society and its participating institutes, the Max Planck Institute for Astronomy, Heidelberg and the Max Planck Institute for Extraterrestrial Physics, Garching, The Johns Hopkins University, Durham University, the University of Edinburgh, Queen's University Belfast, the Harvard-Smithsonian Center for Astrophysics, the Las Cumbres Observatory Global Telescope Network Incorporated, the National Central University of Taiwan, the Space Telescope Science Institute, the National Aeronautics and Space Administration under Grant No.\ NNX08AR22G issued through the Planetary Science Division of the NASA Science Mission Directorate, the National Science Foundation under Grant No.\ AST-1238877, the University of Maryland, and Eotvos Lorand University (ELTE). This publication makes use of data products from the Wide-field Infrared Survey Explorer, which is a joint project of the University of California, Los Angeles, and the Jet Propulsion Laboratory/California Institute of Technology, funded by the National Aeronautics and Space Administration. The CSS survey is funded by the National Aeronautics and Space Administration under Grant No.\ NNG05GF22G issued through the Science Mission Directorate Near-Earth Objects Observations Program. The CRTS survey is supported by the U.S.~National Science Foundation under grants AST-0909182 and AST-1313422.
\end{acknowledgements}

\noindent\flushcolsend
\begin{appendix}
\renewcommand{\thefigure}{A.\arabic{figure}}
\renewcommand{\thetable}{A.\arabic{table}}
\renewcommand{\theequation}{A.\arabic{equation}}
\begin{figure*}
\centering
\includegraphics[height=1\textwidth, angle=-90]{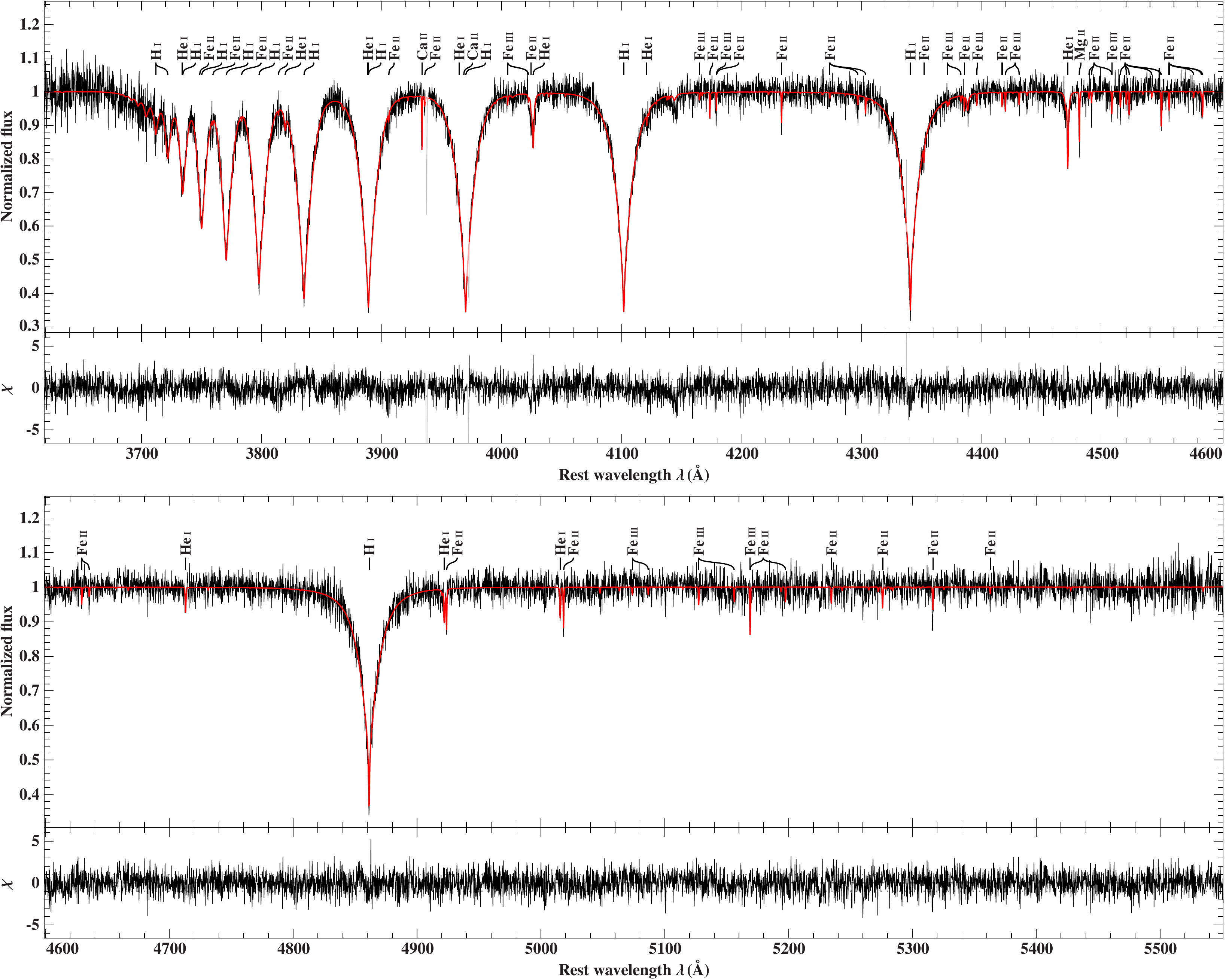}
\caption{\label{fig:spectra}Exemplary comparison of the best fitting model spectrum (red line) with normalized observation (black line; X-shooter spectrum taken on March 23, 2019, i.e., 2\,458\,565.8693\,HJD). Light colors mark regions that have been excluded from fitting, e.g., due to the presence of interstellar or telluric lines. Residuals, $\chi$, are shown as well. The optical spectrum only exhibits lines of H, He, Mg, Ca, and Fe.}
\end{figure*}
\begin{figure*}\ContinuedFloat
\centering
\includegraphics[height=1\textwidth, angle=-90]{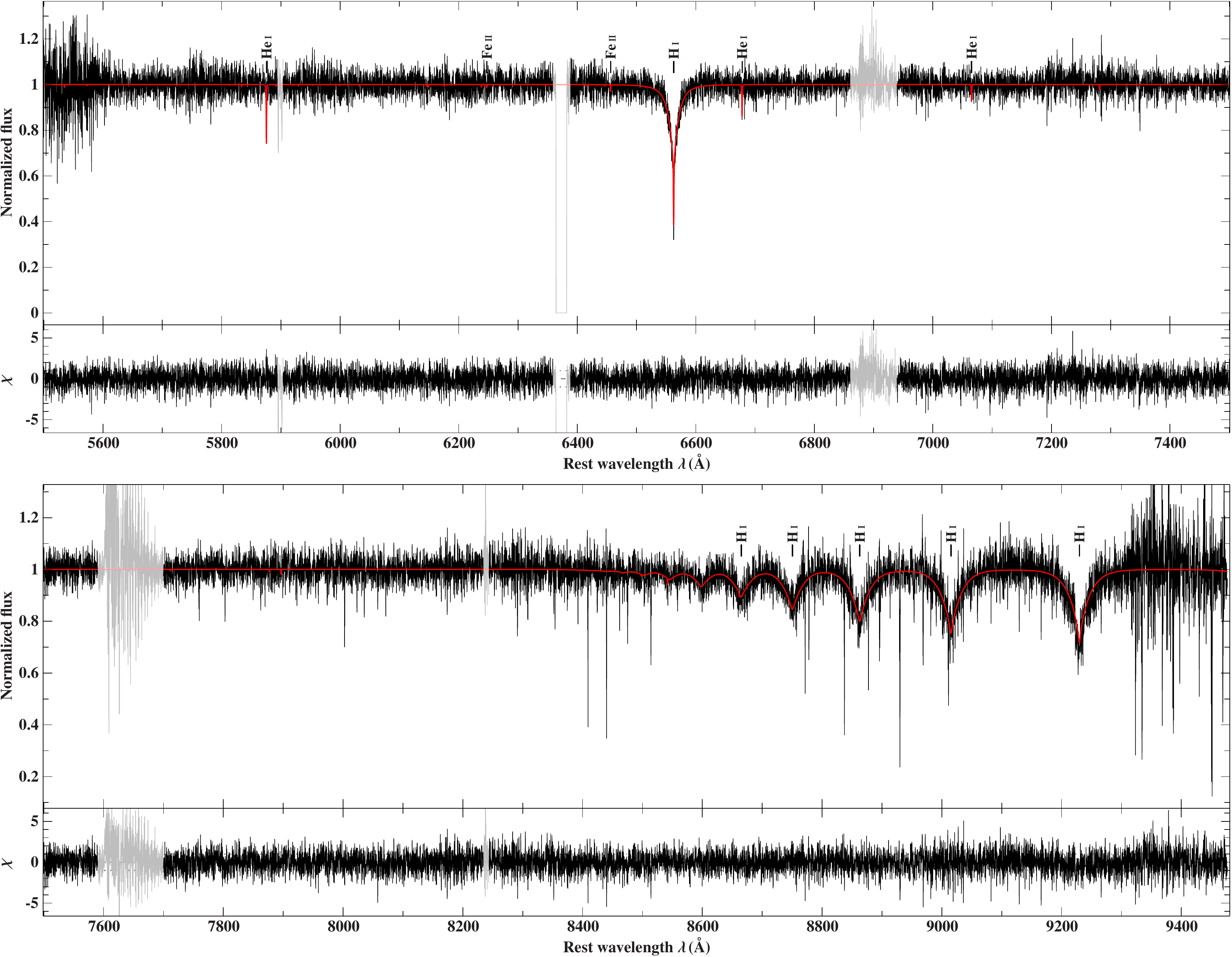}
\caption{Continued.}
\end{figure*}
\begin{figure*}
\centering
\includegraphics[width=1\textwidth]{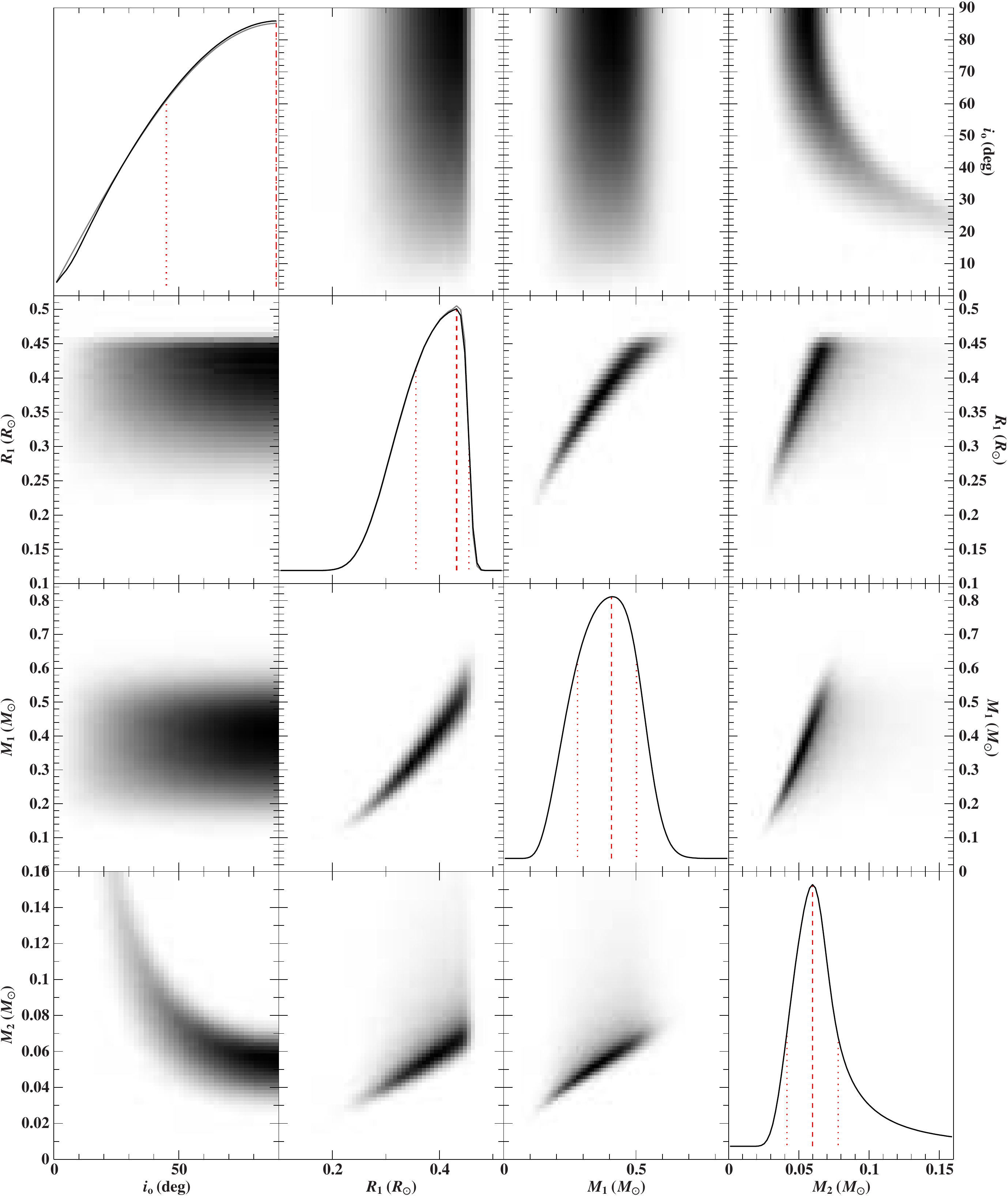}
\caption{\label{fig:corner_plots_Bayes}Corner plot visualizing the correlations between the orbital inclination ($i_\mathrm{o}$), the radius and mass of the visible component ($R_1$, $M_1$), and the mass of its unseen companion ($M_2$) for an RL filling factor $x=1$ (see Eq.~(\ref{eq:condition_h})). The diagonal panels show the respective posterior PDFs. Maximum values are marked by dashed red lines, and lower and upper bounds of the highest density interval are shown as dotted red lines. Prior PDFs for $i_\mathrm{o}$ and $R_1$ according to Eqs.~(\ref{eq:prior_io}) and (\ref{eq:prior_R1}) are drawn as solid gray lines.}
\end{figure*}
\begin{figure*}\ContinuedFloat
\centering
\includegraphics[width=1\textwidth]{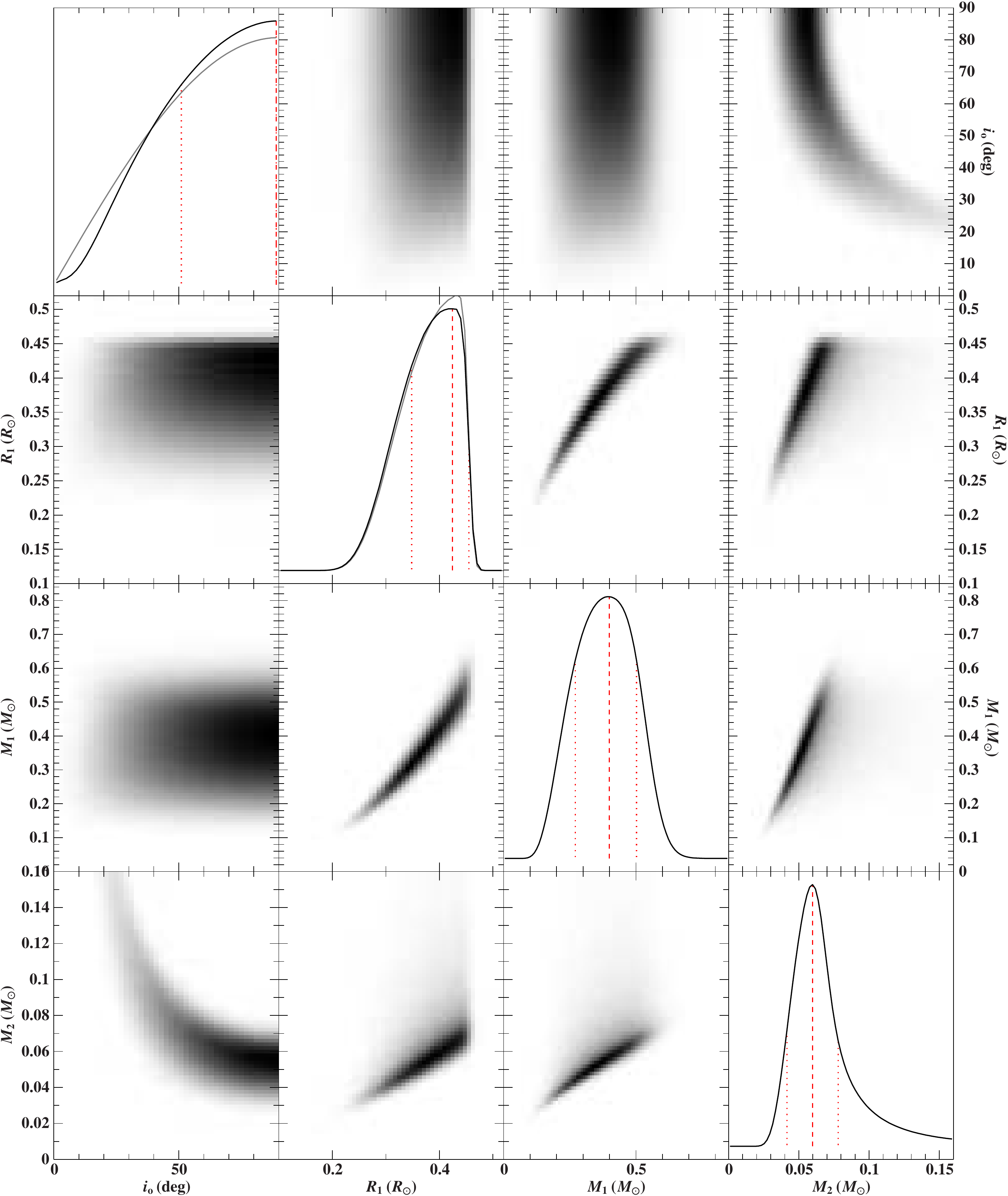}
\caption{Continued for $x=0.9$.}
\end{figure*}
\begin{figure*}\ContinuedFloat
\centering
\includegraphics[width=1\textwidth]{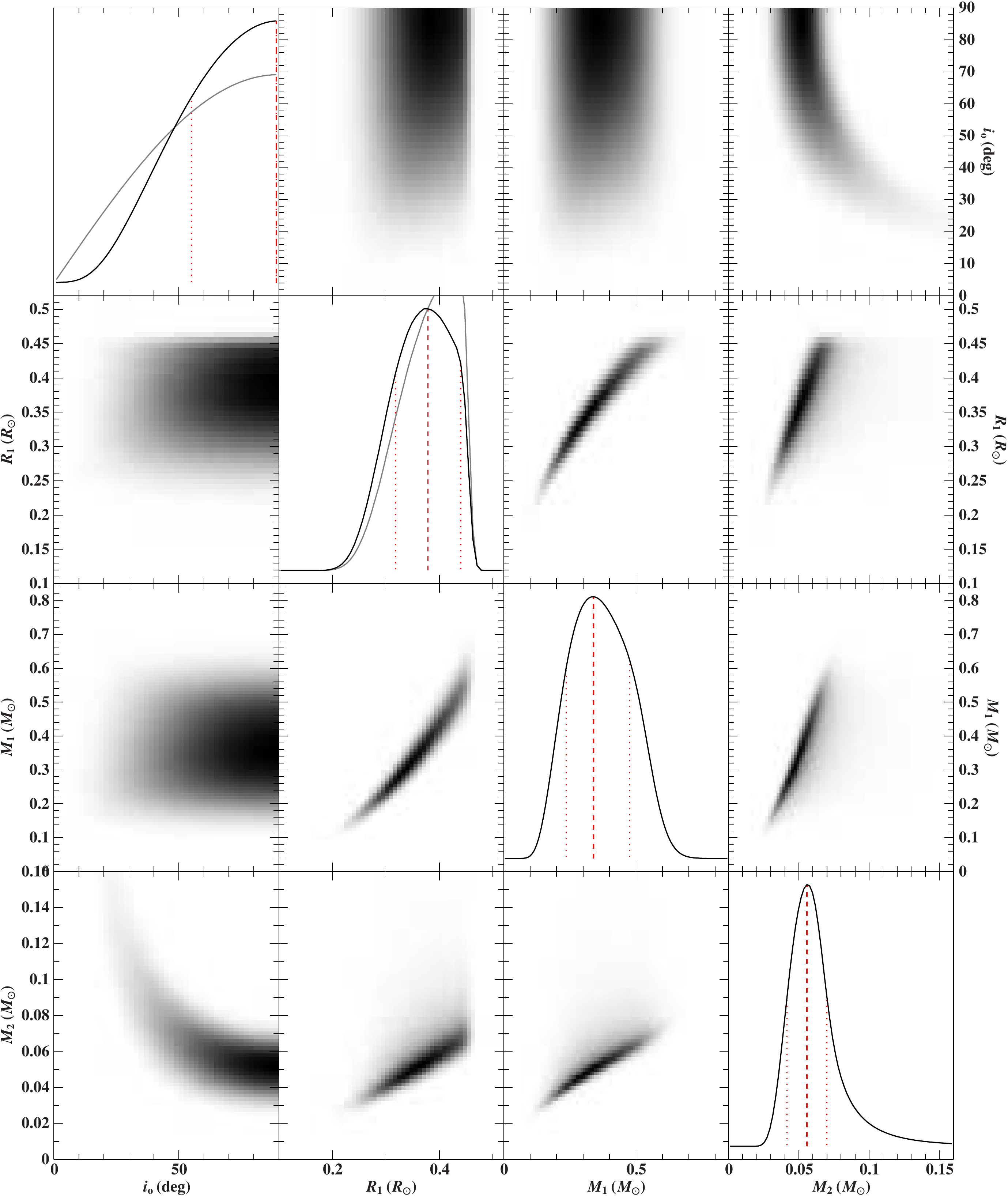}
\caption{Continued for $x=0.8$.}
\end{figure*}
\begin{figure*}\ContinuedFloat
\centering
\includegraphics[width=1\textwidth]{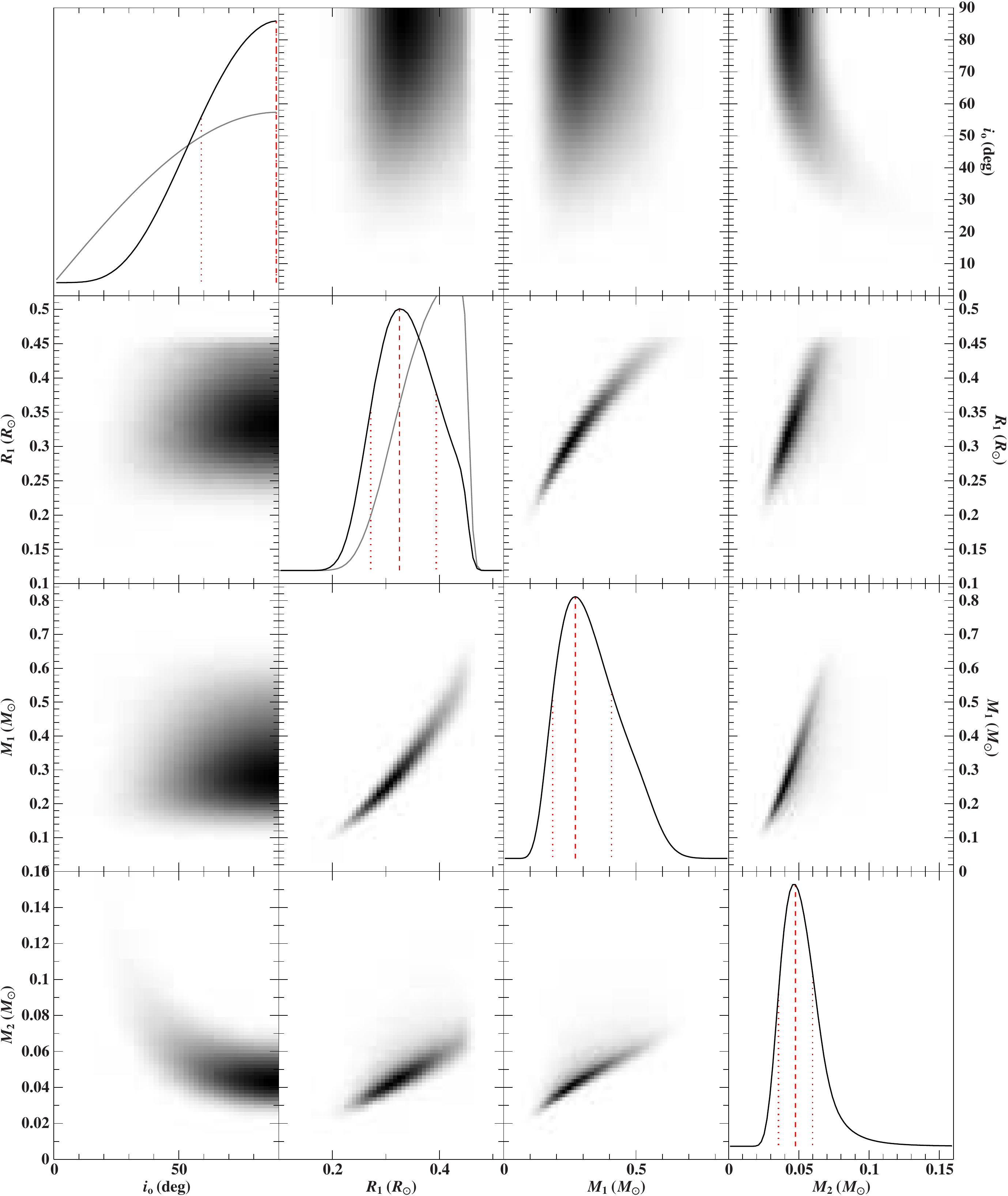}
\caption{Continued for $x=0.7$.}
\end{figure*}
\begin{figure*}\ContinuedFloat
\centering
\includegraphics[width=1\textwidth]{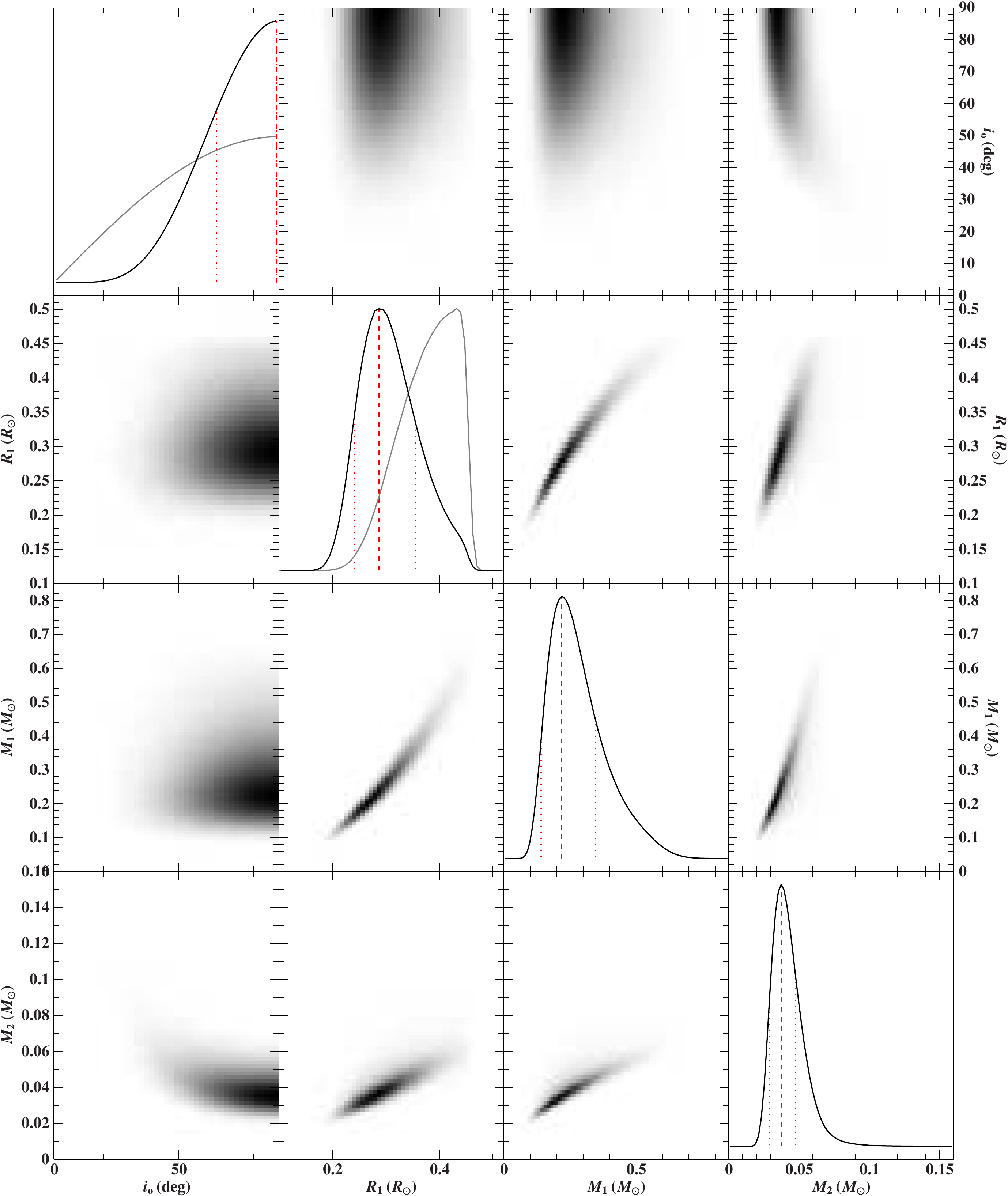}
\caption{Continued for $x=0.6$.}
\end{figure*}
\begin{figure*}\ContinuedFloat
\centering
\includegraphics[width=1\textwidth]{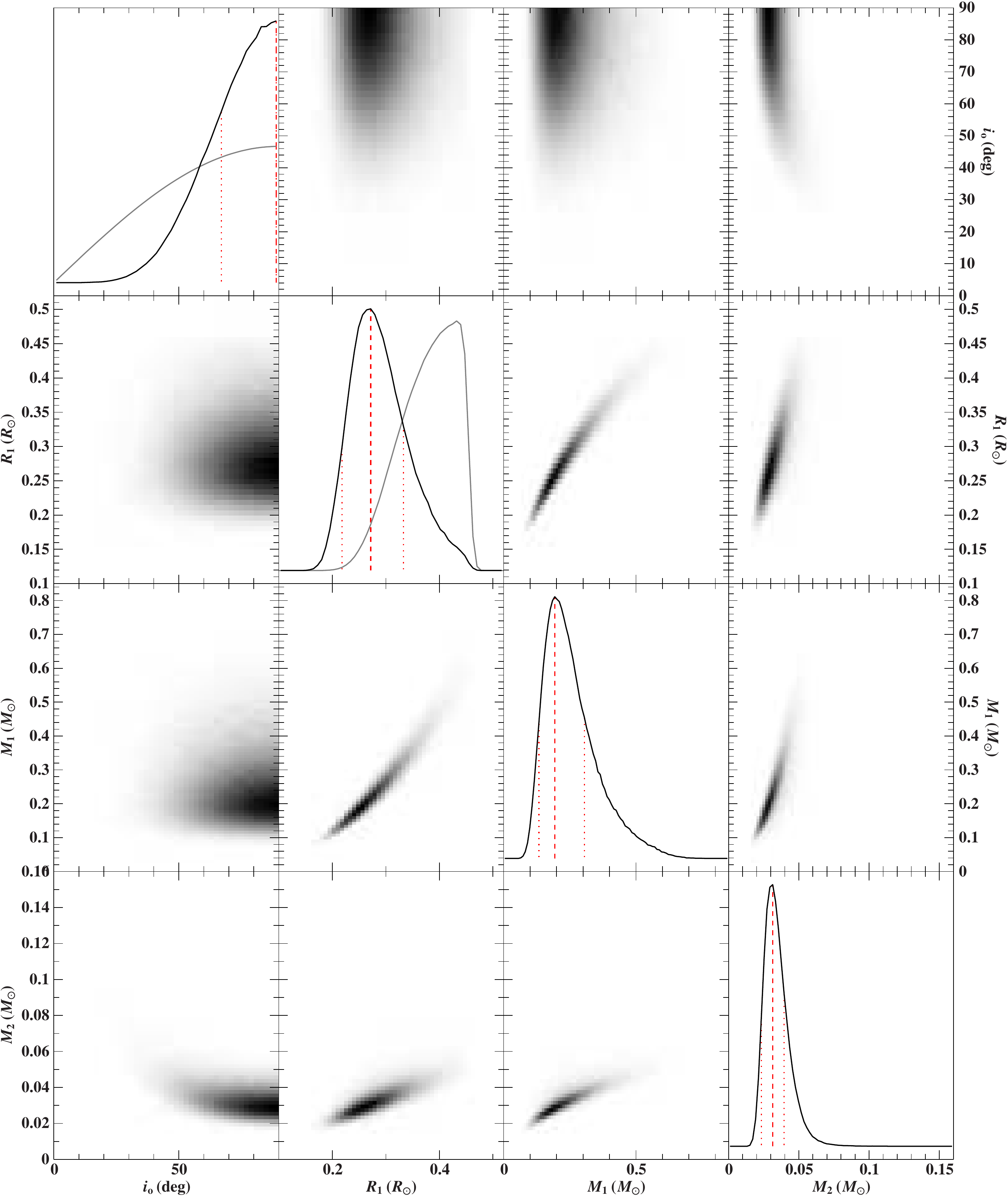}
\caption{Continued for $x=0.5$.}
\end{figure*}
\end{appendix}
\end{document}